%% file: InstabMode13GcVarPart1.tex
\documentclass{elsart}
\usepackage{natbib}
\usepackage{epsfig}
\usepackage{amssymb}
\usepackage{amsmath}
\usepackage[center,small]{subfigure}
\usepackage{color}
\newcommand{\ds}{\displaystyle}

\begin{document}
\begin{frontmatter}

\title{Configurational stability of a crack propagating in a material with mode-dependent fracture energy - Part I: Mixed-mode I+III}

\author{Jean-Baptiste Leblond\corauthref{cor}$^1$},
\author{Alain Karma\corauthref{cor}$^2$},
\corauth[cor]{Corresponding author.}
\author{Laurent Ponson$^1$},
\author{Aditya Vasudevan$^{1,2}$}

\address{$^1$Sorbonne Universit\'{e}, Facult\'{e} des Sciences et Ing\'{e}nierie, Campus Pierre et Marie Curie, CNRS, UMR 7190, Institut Jean Le Rond d'Alembert, F-75005 Paris, France}
\address{$^2$Physics Department and Center for Interdisciplinary Research on Complex Systems, Northeastern University, Boston, MA 02115, USA}

\begin{abstract}

In a previous paper \citep{LKL11}, we proposed a theoretical interpretation of the experimentally well-known instability of coplanar crack propagation in mode I+III. The interpretation relied on a stability analysis based on analytical expressions of the stress intensity factors for a crack slightly perturbed both within and out of its original plane, due to \cite{GR86} and \cite{MGW98}, coupled with a double propagation criterion combining \cite{G20}'s energetic condition and \cite{GS74}'s principle of local symmetry. Under such assumptions instability modes were indeed evidenced for values of the mode mixity ratio - ratio of the mode III to mode I stress intensity factors applied remotely - larger than some threshold depending only on Poisson's ratio.
Unfortunately, the predicted thresholds were much larger than those generally observed for typical values of this material parameter.
While the subcritical character of the nonlinear bifurcation from coplanar to fragmented fronts has been proposed as a possible explanation for this discrepancy \citep{CCLNPK15}, we propose here an alternative explanation based on the introduction of a constitutive relationship between the fracture energy and the mode mixity ratio, which is motivated by experimental observations. By re-examining the linear stability analysis of a planar propagating front, we show that such a relationship suffices, provided that it is strong enough, to lower significantly the threshold value of the mode mixity ratio for instability so as to bring it in a range more consistent with experiments.
Interesting formulae are also derived for the distributions of the perturbed stress intensity factors and energy-release-rate, in the special case of perturbations of the crack surface and front obeying the principle of local symmetry and having reached a stationary state (corresponding to instability modes in near-threshold conditions).

\end{abstract}

\end{frontmatter}


\section{Introduction}\label{sec:Intro}

Exploring the {\it configurational stability} of cracks loaded under mixed-mode conditions is of major interest, as it controls not only the patterns produced by fracture but also the resistance of the material to crack propagation. It amounts to studying the {\it stability of the geometrical configuration} of a crack as it propagates under some arbitrary combination of modes. (The problem of {\it configurational stability} thus defined should not be confused with that of {\it propagation stability}, consisting of the alternative between arrest and acceleration of a crack propagating under constant loading).

The instability of the coplanar configuration of a crack propagating under mixed mode I+III conditions is well-documented experimentally; see the works of \cite{S69,K70,PK75,HP79,PSD82,ST87,PA88,YM89,H93,H95,CP96,L97,LLM01b,LBFW08,LMR10,GO12,PR14}, to quote just a few. Cracks loaded in mode I+III are known to propagate in the form of small facets inclined over the average plane of propagation. Materials as diverse as glass \citep{S69,PR14}, steels \citep{HP79,YM89,L97}, rocks \citep{PSD82,PA88,CP96}, alumina \citep{ST87}, PMMA \citep{LBFW08}, gypsum and cheese (!) \citep{GO12} have been considered in experiments; this brings strength to the idea that the causes of the fragmentation of the crack front are not rooted in microscopic, material-dependent mechanisms, so that the standard macroscopic theory of Linear Elastic Fracture Mechanics (LEFM) should be able to handle the problem. But the breaking of translational invariance in the direction of the crack front resulting from the formation of facets, and the very complex, fully 3D associated crack geometry, greatly complicate the analysis.

Following a long period during which theoretical analyses of the problem remained scarce and incomplete (the works of \cite{L97} and \cite{LLM01a,LLM01b} are typical examples), a more satisfactory - though still imperfect - approach was proposed by \cite{LKL11}, drawing inspiration from the results of numerical simulations of \cite{PK10} based on a phase-field model of \cite{KKL01}. The work consisted of a rigorous linear stability analysis of coplanar propagation of the crack under mixed mode I+III conditions, which relied on two basic elements:
\begin{itemize}
  \item rigorous first-order expressions of the stress intensity factors (SIFs) along the front of a crack slightly perturbed both within and out of its original plane, due to \cite{GR86} for the in-plane perturbation of the crack front, and \cite{MGW98} for the out-of-plane perturbation of the crack surface;
  \item use of a double propagation criterion combining \cite{G20}'s energetic condition and \cite{GS74}'s Principle of Local Symmetry (PLS).
\end{itemize}
Instability modes - perturbed configurations of the crack surface and front growing without bound in time while continuously satisfying the double criterion at all instants and positions along the crack front - were found for values of the mode mixity ratio $\rho^0$ - ratio of the unperturbed mode III SIF $K_{III}^0$ to the unperturbed mode I SIF $K_I^0$ - larger than a threshold value $\rho^{\rm cr}$ given by
\begin{equation}\label{eqn:CritThresh}
  \rho^{\rm cr} = \left[ \frac{(1-\nu)(2-3\nu)}{3(2-\nu)-4\sqrt{2}(1-2\nu)} \right]^{1/2}
\end{equation}
where $\nu$ denotes Poisson's ratio. This quantity
is in the range of $0.4$ to $0.5$ for typical values of $\nu$ for glass and polymers.
Unfortunately it is much larger than most threshold values of the mode mixity ratio actually observed. Although an exceptionally high value of $0.57$ has been reported by \cite{ERK17} in some aluminium alloy, most of the time small or very small thresholds have been observed: for instance \cite{S69}'s estimate for glass amounted to a few percent, and \cite{PR14} even claimed that there was no threshold at all in both Homalite 100 and glass.

Several tentative explanations of this discrepancy have been proposed. A natural idea is that imperfections of the crack front geometry and/or loading could promote deviations from the fundamental coplanar solution below the threshold. This idea was first put to test by \cite{LL15} using \cite{CR80}'s heuristic 2D directional stability criterion, suitably extended to the 3D case; it was shown that small accidental undulations of the crack front within the crack plane, due for instance to small heterogeneities of toughness, may generate a positive non-singular stress in the direction of propagation, thus causing the crack to deviate from coplanarity.
A more complete investigation of the idea was provided by \cite{CCLNPK15} in the form of an extensive numerical study of non-coplanar solutions using \cite{KKL01}'s phase-field model, which has been shown by \cite{HK09} to be equivalent to combining the Griffith condition and PLS in isotropic media.
This study demonstrated that the bifurcation accompanying the transition from coplanar to fragmented front is strongly subcritical, suggesting that large enough perturbations may allow the system to jump from the stable branch to the unstable one well below the theoretical threshold.
\begin{figure}[h]
\centerline{\psfig{figure=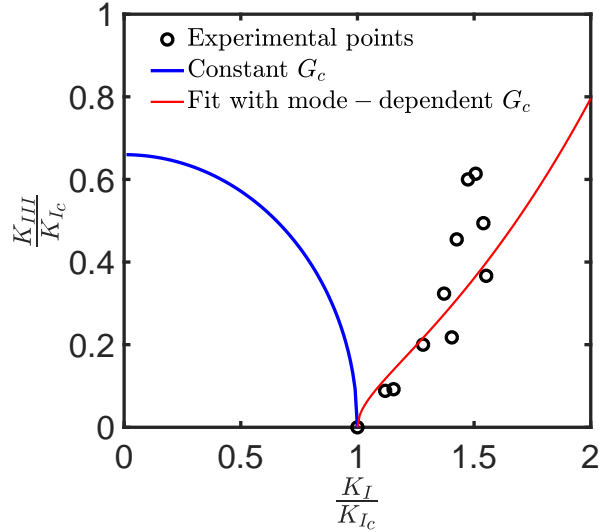,height=7cm}} \caption{Critical stress intensity factors for crack initiation under mixed mode I + III. Black dots: experimental data of \cite{LMR10}; blue line: crack propagation boundary corresponding to a constant fracture energy $G_c$; red line: best fit (for small $\rho$) of the experimental data obtained by assuming a $\rho$-dependent $G_c$ given by equation (\ref{eqn:GcVsRho}) with $\kappa=2$ and $ \gamma \approx 25$.}
\label{fig:KIIIbyKIcVsKIbyKIc}
\end{figure}

The essential aim of this paper is to explore still another possibility, namely the fact that the critical-energy-release rate $G_c$ might not be a constant but depend upon the mode mixity ratio $\rho\equiv K_{III}/K_I$. The dependence of the fracture energy upon the ratio $K_{II}/K_I$ is well documented for interfacial fracture between two distinct materials~\citep{FS03}; this effect plays an important role on the type of fracture patterns emerging from the delamination of thin films from a rigid substrate~\citep{FGBP17}. For bulk fracture under mixed-mode I+III, such an assumption is suggested by the general observation that crack propagation is more difficult under such conditions than in pure mode I. The effect was for instance quantified by \cite{LMR10}; Fig. \ref{fig:KIIIbyKIcVsKIbyKIc} shows these authors' experimental measurement of the critical stress intensity factors at initiation under mode I + III conditions. One clearly sees that the fracture energy required to propagate a crack is higher in mode I + III than in pure mode I.
The assumption of a mode-dependent $G_c$ is also supported indirectly by the fact that as mentioned above, widely different thresholds have been observed in different materials; the modest variations of Poisson's ratio from one material to another do not seem capable of explaining such large variations, which suggests that the threshold may depend on additional material parameters.

The dependence of the critical energy-release-rate $G_c$ upon $\rho = K_{III}/K_I$, just like its dependence upon $K_{II}/K_I$ in interfacial fracture, might arise from the influence of shear on the damage mechanisms occurring within the process zone, at scales disregarded by LEFM. It could also originate from shear-dependent mechanisms taking place at a scale larger than that of the process zone, though still smaller than the ``local'' scale of the stability analysis. For example, possible formation under mode I+III conditions of tilted facets at a sub-continuum-mechanics scale could be invoked, a view suggested by the observation of self-similar fragmentation patterns consisting of facets at various length scales along the crack front reported by \cite{PR14} and also apparent in \cite{CCLNPK15}'s experimental observations.
The aim of this paper is not to study the physical origin of the dependence of $G_c$ on $\rho$, but to assume its existence as a {\it purely heuristic hypothesis}, and explore its consequences on linear stability. The physical intuitive idea is that a shear-dependent fracture energy may promote fragmentation, since fragmentation may result in smaller local values of $K_{III}$ and consequently less energy required for fracture.

Furthermore, to link our theoretical predictions to experiments, we use the experimental data of \cite{LMR10} to estimate the strength of the dependence of $G_c$ on $\rho$ at the local scale as shown in Fig. \ref{fig:KIIIbyKIcVsKIbyKIc}. This procedure is only approximate given that those measurements are made on a crack front that is already fragmented, at least on some small initial scale, before actual propagation of the crack. However, a possible justification of it lies in the aforementioned possibility that the $G_c(\rho)$ relationship may originate from some shear-dependent mechanism operating at scales larger than that of the process zone but smaller than that of the instability wavelength, assumed to set a larger facet scale for propagating fragmented fracture fronts. This hypothesized separation of scales between the mechanism underlying the $G_c(\rho)$ relationship and the initial front fragmentation scale induced by mode I+III crack propagation would presumably continue to hold during the subsequent facet coarsening process that further increases the fragmentation scale \citep{CCLNPK15,LKL15}. However facet coarsening is not investigated in the present study, limited to a linear stability analysis applicable only to the beginning of the development of unstability modes.

The paper is organized as follows:
\begin{itemize}
  \item Section \ref{sec:PertCrack} presents general hypotheses and notations, and formulae - serving as a basis for the analyses to follow - for the SIFs along the front of a slightly perturbed semi-infinite crack, due to \cite{GR86} for the in-plane perturbation of the crack front, and \cite{MGW98} for the out-of-plane perturbation of the crack surface.
  \item Then Section \ref{sec:Mode13StabAnal} extends the stability analysis of \cite{LKL11} of coplanar crack propagation in mixed-mode I+III by including a dependence of the fracture energy upon mode mixity, and examines the impact of this dependence upon the threshold value of the mode mixity ratio.
  \item As a complement, Section \ref{sec:PertStat} investigates some remarkable properties of the perturbed SIFs and energy-release-rate along the crack front in the special case of stationary perturbations obeying the PLS, corresponding to instability modes in near-threshold conditions.
\end{itemize}


\section{First-order perturbation of a semi-infinite crack in an infinite body}\label{sec:PertCrack}

\begin{figure}[h]
\centerline{\psfig{figure=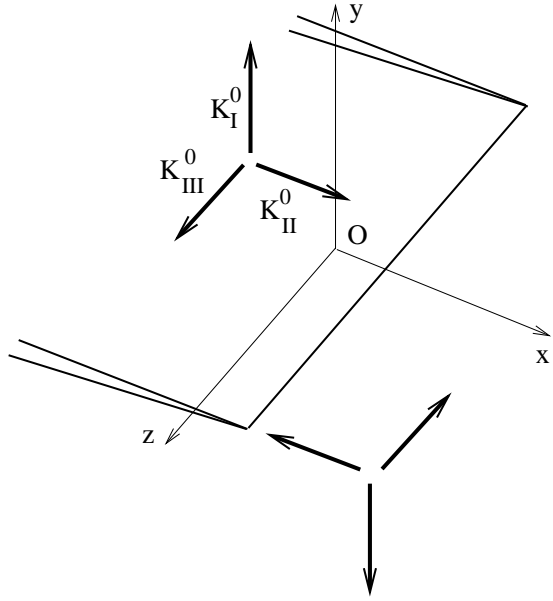,height=8cm}} \caption{Unperturbed geometry and loading.}
\label{fig:InitConfig}
\end{figure}

We consider an infinite body made of some isotropic elastic material and containing a semi-infinite crack. In the initial, unperturbed configuration (Fig. \ref{fig:InitConfig}), this crack is planar and its front is straight. The usual convention for axes is used: the origin $O$ is chosen arbitrarily within the crack plane, the axis $Ox$ oriented along the direction of propagation, the axis $Oy$ perpendicular to the crack plane and the axis $Oz$ parallel to the crack front. The crack is loaded in general mixed-mode I+II+III conditions through some system of forces independent of the coordinate $z$, so that the unperturbed SIFs $K_I^0$, $K_{II}^0$, $K_{III}^0$ are independent of the position of their point of observation along the crack front.

\begin{figure}[h]
  \centering
     \subfigure[\label{fig:InPlanePert} In-plane perturbation of the crack front.]
       {
         \epsfig{figure=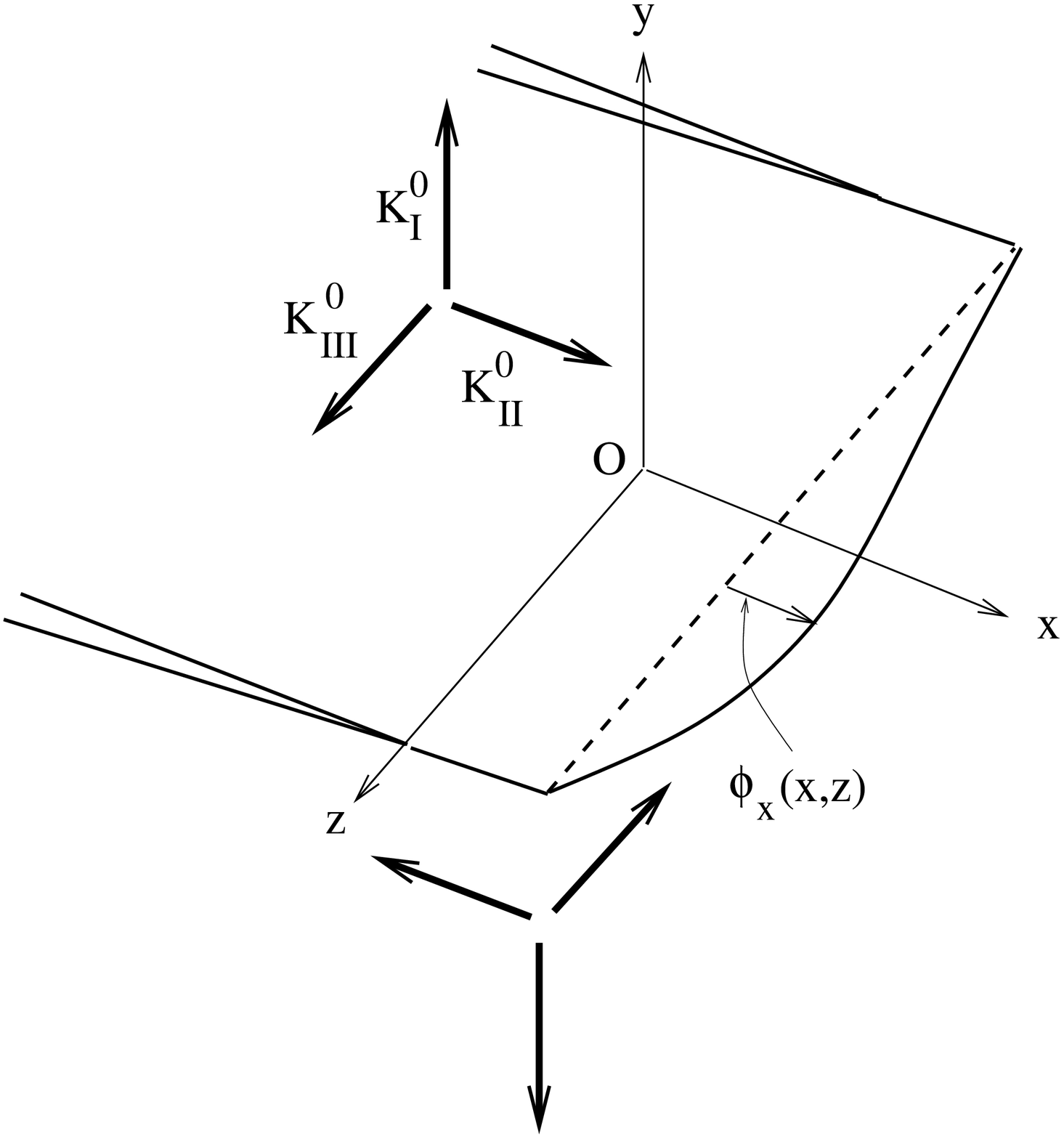,height=7cm}
       }
     \subfigure[\label{fig:OutOfPlanePert} Out-of-plane perturbation of the crack surface.]
       {
         \epsfig{figure=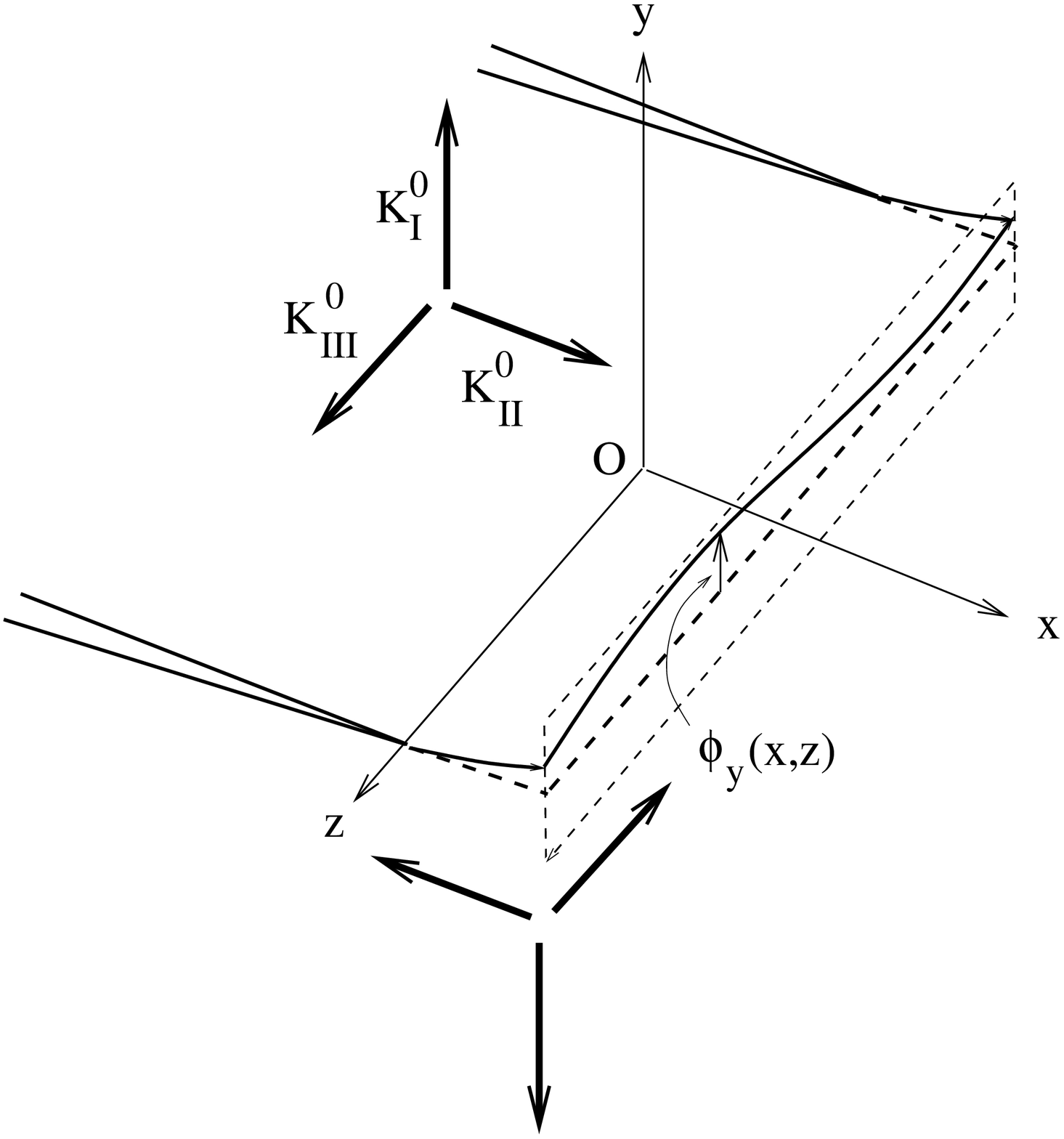,height=7cm}
       }
     \caption{\label{fig:PertConfig} In-plane and out-of-plane perturbations of the crack front and surface.}
\end{figure}

In the perturbed configuration (Fig. \ref{fig:PertConfig}), the crack front is perturbed within the original crack plane by a small amount $\phi_x(x,z)$ in the direction $Ox$ (Fig. \ref{fig:InPlanePert}), and the crack surface is perturbed out of the original crack plane by a small amount $\phi_y(x,z)$ in the direction $Oy$ (Fig. \ref{fig:OutOfPlanePert}); the argument $x$ in the functions here represents the average position of the perturbed crack front in the direction of propagation. (Note that in Fig. \ref{fig:OutOfPlanePert} the perturbation is limited to the immediate vicinity of the crack front for legibility but can in fact extend over the entire crack surface). The perturbation of the $p$-th SIF ($p=I,II,III$) resulting from the in-plane perturbation $\phi_x(x,z)$ of the crack front is denoted $\delta_{x}K_{p}(x,z)$, and that resulting from the out-of-plane perturbation $\phi_y(x,z)$ of the crack surface, $\delta_{y}K_{p}(x,z)$. To first order in the pair $(\phi_x,\,\phi_y)$, these perturbations are additive so that the total perturbation of the $p$-th SIF is simply
\begin{equation}\label{eqn:AddPertSIF}
  \delta K_{p}(x,z) = \delta_{x}K_{p}(x,z) + \delta_{y}K_{p}(x,z) \quad (p=I,II,III).
\end{equation}

We shall apply the results of \cite{GR86} on the in-plane perturbations of the SIFs, $\delta_{x}K_{p}(x,z)$, and those of \cite{MGW98} on their out-of-plane perturbations, $\delta_{y}K_{p}(x,z)$. To do so, we introduce the assumption that the characteristic length defined by the loading, in the absence of any length scale defined by the infinite geometry itself, is much larger than the typical distances of variation of the perturbations $\phi_x(x,z)$ and $\phi_y(x,z)$ of the crack front and surface. It is then possible to retain only these terms, in \cite{GR86}'s and \cite{MGW98}'s formulae, which involve the unperturbed SIFs, discarding those involving the unperturbed non-singular stresses and higher-order constants characterizing the initial near-front stress field. \cite{GR86}'s formulae then read
\begin{equation}\label{eqn:dxK}
  \left\{
    \begin{array}{lll}
      \ds \delta_xK_{I}(x,z) & = & \ds \frac{K_I^0}{2\pi} \, PV \int_{-\infty}^{+\infty}\frac{\phi_x(x,z')-\phi_x(x,z)}{(z'-z)^2}\,dz'  \\[3.mm]
      \ds \delta_xK_{II}(x,z) & = & \ds - \frac{2}{2-\nu} K_{III}^0\,\frac{\partial \phi_x}{\partial z}(x,z)
      + \frac{2-3\nu}{2-\nu}\,\frac{K_{II}^0}{2\pi} \, PV \int_{-\infty}^{+\infty}\frac{\phi_x(x,z')-\phi_x(x,z)}{(z'-z)^2}\,dz' \\[3.mm]
      \ds \delta_xK_{III}(x,z) & = & \ds \frac{2(1-\nu)}{2-\nu} K_{II}^0\frac{\partial \phi_x}{\partial z}(x,z) + \frac{2+\nu}{2-\nu}\,\frac{K_{III}^0}{2\pi} \, PV \int_{-\infty}^{+\infty}\frac{\phi_x(x,z')-\phi_x(x,z)}{(z'-z)^2}\,dz'  \\
    \end{array}
  \right.
\end{equation}
where the symbol $PV$ denotes a Cauchy Principal Value; and \cite{MGW98}'s formulae read
\begin{equation}\label{eqn:dyK}
  \left\{
    \begin{array}{lll}
      \ds \delta_yK_I(x,z) & = & \ds - \frac{3}{2}K_{II}^0\frac{\partial \phi_y}{\partial x}(x,z) - 2K_{III}^0\,\frac{\partial \phi_y}{\partial z}(x,z) \\[3.mm]
      {}                   & {} & \ds - \frac{K_{II}^0}{2\pi} \, PV \int_{-\infty}^{+\infty}\frac{\phi_y(x,z')-\phi_y(x,z)}{(z'-z)^2}\,dz'
                                      + \delta_y K_I^{\rm skew}(x,z) \\[3.mm]
      \ds \delta_yK_{II}(x,z) & = & \ds \frac{K_{I}^0}{2}\,\frac{\partial \phi_y}{\partial x}(x,z)
                                        - \frac{2-3\nu}{2-\nu}\frac{K_I^0}{2\pi} \, PV \int_{-\infty}^{+\infty}\frac{\phi_y(x,z')-\phi_y(x,z)}{(z'-z)^2}\,dz'  \\[3.mm]
      \ds \delta_yK_{III}(x,z) & = & \ds \frac{2(1-\nu)^2}{2-\nu}K_{I}^0\,\frac{\partial \phi_y}{\partial z}(x,z). \\
    \end{array}
  \right.
\end{equation}
In the expression of $\delta_yK_I(x,z)$ here the quantity $\delta_y K_I^{\rm skew}(x,z)$ is connected to \cite{B87}'s {\it skew-symmetric} crack-face weight functions - whence the notation. \cite{MGW98}'s calculation of this term for a perturbation $\phi_y(x,z)$ independent of $x$ has been extended by \cite{LKL11} to an arbitrary perturbation, with the following result:
\begin{equation}\label{eqn:dyK1skew}
  \delta_y K_I^{\rm skew}(x,z) = \frac{\sqrt{2}}{4\pi} \, \frac{1-2\nu}{1-\nu} \, {\rm Re} \left\{\int_{-\infty}^{x} dx' \int_{-\infty}^{+\infty}
                               \frac{[K_{III}^0-i(1-\nu)K_{II}^0](\partial \phi_y/\partial z)(x',z')}{(x-x')^{1/2}\left[x-x'+i(z-z')\right]^{3/2}} \, dz' \right\}
\end{equation}
where the cut of the complex power function is along the half-line of negative real numbers.


\section{Linear stability analysis}\label{sec:Mode13StabAnal}

\subsection{Generalities}\label{subsec:Mode13Gen}

From now on the unperturbed SIF $K_{II}^0$ of mode II will be assumed to be zero so that the crack will be loaded in mixed-mode I+III. Without any loss of generality, we may assume the unperturbed SIF $K_{III}^0$ of mode III to be positive like that of mode I, $K_{I}^0$.

Like in our previous work \citep{LKL11}, the prediction of the crack path will be based on a double propagation criterion enforced at all points of the crack front and all instants, consisting of
\begin{itemize}
  \item \cite{G20}'s condition $G(x,z)=G_c(x,z)$ where $G(x,z)$ denotes the local energy-release-rate, and $G_c(x,z)$ the local critical value of this quantity;
  \item \cite{GS74}'s PLS stipulating that the local SIF $K_{II}(x,z)$ of mode II must be zero.
\end{itemize}
However, whereas the critical energy-release-rate $G_c$ was supposed to be a constant in the previous work, it will be assumed here to be a function of the mode mixity ratio $\rho$ (ratio of the local mode III to mode I SIFs):
\begin{equation}\label{eqn:DefRho}
  G_c(x,z) \equiv G_c[\rho(x,z)] \quad , \quad \rho(x,z) \equiv \frac{K_{III}(x,z)}{K_{I}(x,z)} \quad (>0).
\end{equation}

Special attention will be paid to the case where the function $G_c(\rho)$ is of the form
\begin{equation}\label{eqn:GcVsRho}
  G_c(\rho) \equiv G_{cI}(1 + \gamma \rho^{\kappa})
\end{equation}
where $G_{cI}$ denotes the value of $G_c$ in pure mode I ($\rho=0$), and $\gamma$ and $\kappa$ dimensionless material parameters.\footnote{Note that formula (\ref{eqn:GcVsRho}) is given only for positive values of the mode mixity ratio $\rho$; for arbitrary values it should be applied with $|\rho|$ instead of $\rho$, the function $G_c(\rho)$ being even for obvious symmetry reasons.} These parameters are assumed to be positive, meaning that the presence of mode III is assumed to {\it increase} the value of the critical energy-release-rate, in line with \cite{FS03}'s observation that the presence of shear systematically results in an increase of the cracking resistance of interfaces between dissimilar materials. A quadratic variation of the fracture energy with mode mixity - $\kappa = 2$ - would be the most natural assumption, as it corresponds to the leading term in a development of the even function $G_c(\rho)$ for small values of $\rho$; such a value is generally chosen to describe the variations of $G_c$ with the ratio $K_{II}/K_{I}$ in interfacial fracture, see e.g.~\cite{FGBP17}. But we choose here to keep $\kappa$ as a free parameter so as not to restrict the generality of the stability analysis.

The analysis will be based on consideration of instability modes consisting of in-plane perturbations of the crack front and out-of-plane perturbations of the crack surface of the form
\begin{equation}\label{eqn:InstMode}
  \left\{
    \begin{array}{lll}
      \phi_x(x,z) & = & e^{\lambda x}\,\psi_x(z) \\
      \phi_y(x,z) & = & e^{\lambda x}\,\psi_y(z)
    \end{array}
  \right.
\end{equation}
where $\lambda$ is a positive parameter homogeneous to the inverse of a length, characterizing the growth rate of the mode in the direction of propagation, and $\psi_x(z)$, $\psi_y(z)$ functions to be determined. (The case $\lambda <0$ could be studied as well, but is of little interest since it corresponds to banal stability of the coplanar configuration of the propagating crack).

Use will be made of Fourier transforms in the direction $z$ of the crack front; the definition adopted here for the Fourier transform $\widehat{\chi}(k)$ of an arbitrary function $\chi(z)$ is
\begin{equation}\label{eqn:DefFourier}
  \chi(z) = \int_{-\infty}^{+\infty} \widehat{\chi}(k) \,e^{ikz} dk \quad \Leftrightarrow \quad
  \widehat{\chi}(k) = \frac{1}{2\pi} \int_{-\infty}^{+\infty} \chi(z) \,e^{-ikz} dz.
\end{equation}
We then define a dimensionless ``normalized growth rate'' $\xi$ of the instability mode by the formula
\begin{equation}\label{eqn:DefXi}
  \xi \equiv \frac{\lambda}{|k|} \quad (>0);
\end{equation}
this parameter compares the growth rate of the instability mode, $\lambda$, in the direction $x$, to the wavenumber of the Fourier component considered, $|k|$, in the direction $z$.


\subsection{Fourier transforms of the perturbations of the energy-release-rate and the mode mixity ratio}\label{subsec:Mode13FourierDeltaGRho}

To calculate the Fourier transforms $\widehat{\delta G}$, $\widehat{\delta\rho}$ of the first-order variations $\delta G$, $\delta\rho$ of the energy-release-rate and mode mixity ratio, the first task is to calculate the Fourier transforms $\widehat{\delta K_p}$ of the variations of the SIFs $\delta K_p$ ($p=I,II,III$). The simplest way of doing so is to insert the expressions (\ref{eqn:InstMode}) of the perturbations $\phi_x(x,z)$, $\phi_y(x,z)$ into equations (\ref{eqn:AddPertSIF} - \ref{eqn:dyK1skew}), use the expressions (\ref{eqn:DefFourier})$_1$ of the functions $\psi_x(z)$, $\psi_y(z)$ in terms of their Fourier transforms $\widehat{\psi_x}(k)$, $\widehat{\psi_y}(k)$, and evaluate the integrals for each value of $k$. The calculations are elementary except for that of the term $\delta_y K_I^{\rm skew}$, which is presented in Appendix \ref{app:FourierDeltaK1Skew} with the following result:\footnote{In our previous stability analysis \citep{LKL11}, $F$ was considered to be a function of $|k|/\lambda$ instead of $\xi={\lambda}/{|k|}$, and therefore expressed as $\sqrt{\frac{|k|/\lambda}{|k|/\lambda+1}}$ rather than $1/\sqrt{1+{\lambda}/{|k|}}$.}
\begin{equation}\label{eqn:FourierDeltaK1Skew}
  \widehat{\delta_y K_I^{\rm skew}}(x,k) = \frac{i}{\sqrt{2}} \, \frac{1-2\nu}{1-\nu} \, K_{III}^0 \, e^{\lambda x} F(\xi) \, k\widehat{\psi_y}(k)
  \quad {\rm where} \quad
  F(\xi) \equiv \frac{1}{\sqrt{1+\xi}}\ .
\end{equation}
With the result (\ref{eqn:FourierDeltaK1Skew}), the following expressions of the Fourier transforms of the variations $\delta K_p$ ($p=I,II,III$) of the SIFs are obtained:
\begin{equation}\label{eqn:FourierDeltaKp}
  \left\{
    \begin{array}{lll}
      \widehat{\delta K_I}(x,k) & = & \ds e^{\lambda x} \left\{ - K_I^0\frac{|k|}{2} \widehat{\psi_x}(k)
                                        + iK_{III}^0\,\left[ - 2 + \frac{1-2\nu}{\sqrt{2}(1-\nu)}F(\xi) \right] k \widehat{\psi_y}(k) \right\}  \\[3mm]
      \widehat{\delta K_{II}}(x,k) & = & \ds e^{\lambda x} \left\{ - iK_{III}^0\frac{2}{2-\nu} k \widehat{\psi_x}(k)
                                        + K_I^0\left[ \frac{2-3\nu}{2(2-\nu)}|k| + \frac{\lambda}{2} \right] \widehat{\psi_y}(k) \right\}  \\[3mm]
      \widehat{\delta K_{III}}(x,k) & = & \ds e^{\lambda x} \left\{ - K_{III}^0\,\frac{2+\nu}{2(2-\nu)} |k| \widehat{\psi_x}(k)
                                        + iK_I^0 \frac{2(1-\nu)^2}{2-\nu} k \widehat{\psi_y}(k) \right\}.
    \end{array}
  \right.
\end{equation}


\subsection{Application of the principle of local symmetry}\label{subsec:Mode13ApplicPLS}

\cite{GS74}'s PLS, written in ``Fourier's form'' $\widehat{\delta K_{II}}(x,k) = 0$, then yields
\begin{equation}\label{eqn:PhiyVsPhix}
  \widehat{\psi_y}(k) = \frac{4i}{2-3\nu+(2-\nu)\xi}\, \rho^0\,{\rm sgn}(k)\widehat{\psi_x}(k),
\end{equation}
where
\begin{equation}\label{eqn:DefRho0}
  \rho^0 \equiv \frac{K_{III}^0}{K_I^0}
\end{equation}
denotes the unperturbed mode mixity ratio and ${\rm sgn}(x)$ the sign of $x$.


\subsection{Application of \cite{G20}'s criterion}\label{subsec:Mode13ApplicGriffith}

Using then equations (\ref{eqn:FourierDeltaKp}, \ref{eqn:PhiyVsPhix}), Irwin's formula $G=\frac{1-\nu^2}{E}(K_I^2+K_{II}^2)+\frac{1+\nu}{E}K_{III}^2$ ($E$: Young's modulus) and the definition (\ref{eqn:DefRho})$_2$ of the mode mixity ratio, one can get the Fourier transforms $\widehat{\delta G}$, $\widehat{\delta\rho}$ of the perturbations of the energy-release-rate and mode mixity ratio, as functions of the Fourier transform $\widehat{\psi_x}$ of the sole in-plane perturbation of the crack front. The results, obtained after straightforward but heavy calculations, are as follows:
\begin{equation}\label{eqn:DeltaGDeltaRho}
  \left\{
    \begin{array}{lll}
      \ds \frac{\widehat{\delta G}(k)}{G^0} & = & - e^{\lambda x}f(\rho^0;\xi)\,|k|\widehat{\psi_x}(k) \\
      \widehat{\delta\rho}(k) & = & - e^{\lambda x}g(\rho^0;\xi)\,|k|\widehat{\psi_x}(k)
    \end{array}
  \right.
\end{equation}
where
\begin{equation}\label{eqn:DefG0}
  G^0 \equiv \frac{1-\nu^2}{E}(K_I^0)^2 + \frac{1+\nu}{E}(K_{III}^0)^2
\end{equation}
denotes the unperturbed energy-release-rate, and $f$ and $g$ the functions defined by
\begin{equation}\label{eqn:DefFG}
  \left\{
    \begin{array}{lll}
      f(\rho;\xi) & \equiv & \ds \frac{1-\nu}{1-\nu+\rho^2} \left\{ 1 - \frac{3(2-\nu)-4\sqrt{2}(1-2\nu)F(\xi)-(2+\nu)\xi}
                                    {(1-\nu)\left[ 2-3\nu+(2-\nu)\xi \right]}\,\rho^2 \right\} \\[5mm]
      g(\rho;\xi) & \equiv & \ds \frac{\rho}{2-3\nu+(2-\nu)\xi} \left\{ 4-5\nu+\nu\xi
                                    +2\left[ 4-\sqrt{2}\,\frac{1-2\nu}{1-\nu}F(\xi) \right]\rho^2 \right\}
    \end{array}
  \right.
  \quad (\xi>0).
\end{equation}

To enforce \cite{G20}'s criterion $G(x,z)=G_c(x,z)$ at every point of the crack front and every instant, we use the first-order expansions of $G(x,z)$ and $G_c(x,z)$ in the pair $(\phi_x,\phi_y)$,
\begin{equation}\label{eqn:Expans}
  \left\{
    \begin{array}{lll}
      G(x,z) & = & G^0 + \delta G(x,z) \\
      G_c(x,z) & = & \ds G_c(\rho^0) + \frac{dG_c}{d\rho}(\rho^0) \,\delta\rho(x,z)
    \end{array}
  \right.
\end{equation}
where $\delta G$ and $\delta\rho$ are given, in Fourier's form, by equations (\ref{eqn:DeltaGDeltaRho}). Equating successive terms of the expansions of $G(x,z)$ and $G_c(x,z)$, we get:
\begin{itemize}
  \item At order 0:
    \begin{equation}\label{eqn:Ord0}
      G^0 = G_c(\rho^0) \quad (=G_{cI}[1 + \gamma (\rho^0)^{\kappa}]).
    \end{equation}
    This condition specifies the intensity of the loading inducing propagation of the crack, for the given value of the unperturbed mode mixity ratio $\rho^0$.
  \item At order 1:
  \begin{equation*}
    \delta G(x,z) = \frac{dG_c}{d\rho}(\rho^0) \,\delta\rho(x,z) \quad \Rightarrow
  \end{equation*}
  \begin{equation*}
    G^0 e^{\lambda x}f(\rho^0;\xi)\,|k|\widehat{\psi_x}(k) = \frac{dG_c}{d\rho}(\rho^0) \,e^{\lambda x}g(\rho^0;\xi)\,|k|\widehat{\psi_x}(k) \quad (\forall k).
  \end{equation*}
  There are then two possibilities for each value of the wavenumber $k$: either the Fourier component $\widehat{\psi_x}(k)$ is zero and this relation is trivially satisfied; or it is not, and then necessarily
  \begin{equation}\label{eqn:Ord1}
    f(\rho^0;\xi) - \frac{d({\rm ln}G_c)}{d\rho}(\rho^0)\,g(\rho^0;\xi) = 0
  \end{equation}
  where equation (\ref{eqn:Ord0}) has been used. This condition specifies the value of the normalized growth rate $\xi={\lambda}/{|k|}$ of the instability mode as a function of the unperturbed mode mixity ratio $\rho^0=K_{III}^0/K_{I}^0$.
\end{itemize}


\subsection{Growth rate of the instability mode and critical mode mixity ratio}\label{subsec:Mode13GrowthRate}

Equation (\ref{eqn:Ord1}) may be put in an alternative format by ``solving'' it with respect to $\xi$, formally considering the value of $F(\xi)$ as known; the function $G_c(\rho)$ being assumed to be of the form (\ref{eqn:GcVsRho}), we get upon use of equations (\ref{eqn:DefFG}):
\begin{equation}\label{eqn:EqOnXi}
  \xi = \frac{N(\rho^0;\xi)}{D(\rho^0)} \ \ \mbox{where}
  \left\{
    \begin{array}{lll}
      N(\rho;\xi) & \equiv & \ds -(1-\nu)(2-3\nu) + \left[ 3(2-\nu) - 4\sqrt{2}(1-2\nu)F(\xi) \right]\rho^2 \\[3mm]
      {} & {} & \ds + X(\rho) \left( 1-\nu+\rho^2 \right) \left\{ 4-5\nu+2\left[ 4-\sqrt{2}\frac{1-2\nu}{1-\nu}F(\xi) \right]\rho^2 \right\} \\[3mm]
      D(\rho) & \equiv & \ds (1-\nu)(2-\nu) + (2+\nu)\rho^2 - X(\rho)\nu \left( 1-\nu+\rho^2 \right) \\[3mm]
      X(\rho) & \equiv & \ds \frac{\kappa\gamma\rho^{\kappa}}{1+\gamma\rho^{\kappa}}\,.
    \end{array}
  \right.
\end{equation}

The terms $3(2-\nu) - 4\sqrt{2}(1-2\nu)F(\xi)$ and $4-\sqrt{2}\frac{1-2\nu}{1-\nu}F(\xi)$ in the expression of the numerator $N(\rho;\xi)$ here vary modestly with $\xi$,\footnote{For a standard value of $\nu$ of $0.3$, when $\xi$ goes from $0$ to $+\infty$, the first term varies from $2.84$ to $5.1$, and the second from $3.19$ to $4$. In the limit $\nu \rightarrow 1/2$, they both become constant.} so that they may be considered as approximately constant. Thus equation (\ref{eqn:EqOnXi}) ``almost'' directly provides the value of the normalized growth rate $\xi$, which makes it convenient for both qualitative physical analyses of its predictions, and simple methods of numerical solution such as the fixed-point algorithm. The form (\ref{eqn:EqOnXi}) is also convenient for a mathematical analysis of the solutions in $\xi$; such an analysis is presented in Appendix \ref{app:MathStudy}, with the following conclusions:
\begin{enumerate}
  \item {\it The equation $N(\rho^0;0)=0$ admits a unique positive solution $\rho^{\rm cr}$}. This solution represents the critical value of the unperturbed mode mixity ratio for which a neutrally stable mode (having a zero normalized growth rate $\xi$) exists.
  \item Assume that $\kappa \leq 3$. Then {\it for values of the unperturbed mode mixity ratio $\rho^0$ larger than $\rho^{\rm cr}$, the equation $\xi = {N(\rho^0;\xi)}/{D(\rho^0)}$ on $\xi$ admits a unique positive solution}. This means that an instability mode exists, with normalized growth rate given by this solution.
  \item Assume that $\kappa \leq 3$ {\it and} $\nu \geq \frac{2}{31}(9-5\sqrt{2}) \simeq 0.124$.\footnote{This value may not be optimal, since the proof of Appendix \ref{app:MathStudy} is based on strict inequalities which may be liable to improvement.} Then {\it for values of $\rho^0$ smaller than $\rho^{\rm cr}$, the equation $\xi = {N(\rho^0;\xi)}/{D(\rho^0)}$ on $\xi$ does not admit any positive solution}. Hence no instability mode exists.
\end{enumerate}

It is finally worth noting that when the critical energy-release-rate $G_c$ is a constant ($\gamma=0$), equation (\ref{eqn:EqOnXi}) for the normalized growth rate $\xi$ of the instability mode, and the equation $N(\rho^0;0)=0$ defining the critical mode mixity ratio $\rho^{\rm cr}$, are identical to those obtained in our previous stability analysis, equations (20) and (21) of \citep{LKL11}. They are however new in general.


\subsection{Geometry of the instability mode}\label{subsec:Mode13GeomMode}

Assuming now that $\rho^0 \geq \rho^{\rm cr}$ so that equation (\ref{eqn:EqOnXi}) on $\xi$ admits a solution, we discuss the geometrical shape of the corresponding instability mode. Two cases must be distinguished.

\begin{itemize}
  \item {\it Case 1: $\rho^0 > \rho^{\rm cr}$}.\vskip2mm
   As established in Appendix \ref{app:MathStudy}, equation (\ref{eqn:EqOnXi}) admits a single positive solution in $\xi=\frac{\lambda}{|k|}$. For a fixed (positive) value of the growth rate $\lambda$, this value corresponds to a pair of distinct solutions $(k,-k)$ with $k>0$. Then the Fourier transform $\widehat{\psi_x}(k')$ of the in-plane perturbation of the crack front is nonzero only for $k'=k$ or $k'=-k$. Since the perturbation $\psi_x$ is real, $\widehat{\psi_x}(-k') = \overline{\widehat{\psi_x}(k')}$, and it follows that $\widehat{\psi_x}(k')$ must be of the form
      $\frac{1}{2}\left[ C_x \,\delta(k'-k) + \overline{C_x} \,\delta(k'+k) \right]$ where $C_x$ is a complex number and $\delta$ Dirac's function, or equivalently
      \begin{equation}\label{eqn:PsixHat}
        \widehat{\psi_x}(k') \equiv \frac{A_x}{2}\left[ e^{i\theta}\,\delta(k'-k) + e^{-i\theta}\,\delta(k'+k) \right]
      \end{equation}
      where $A_x$ and $\theta$ are real numbers. Then by equation (\ref{eqn:DefFourier})$_1$, the in-plane perturbation $\psi_x$ of the crack front is given by
      \begin{equation}\label{eqn:Psix}
        \psi_x(z) = \int_{-\infty}^{+\infty} \widehat{\psi_x}(k') \,e^{ik'z} dk = \frac{A_x}{2}\left[ e^{i(kz+\theta)} + e^{-i(kz+\theta)}\right]
                  = A_x \cos(kz+\theta).
      \end{equation}
      By equation (\ref{eqn:PhiyVsPhix}), the expression of the Fourier transform $\widehat{\psi_y}(k')$ of the out-of-plane perturbation of the crack surface then reads
      \begin{equation}\label{eqn:PsiyHat}
          \widehat{\psi_y}(k') = \frac{4i}{2-3\nu+(2-\nu)\xi}\, \rho^0\,\frac{A_x}{2}\left[ e^{i\theta}\,\delta(k'-k) - e^{-i\theta}\,\delta(k'+k) \right].
      \end{equation}
      This implies that the out-of-plane perturbation $\psi_y$ of the crack surface is given by
      \begin{equation}\label{eqn:Psiy}
        \begin{array}{lll}
          \psi_y(z) & = & \ds \int_{-\infty}^{+\infty} \widehat{\psi_y}(k') \,e^{ik'z} dk = \frac{4i}{2-3\nu+(2-\nu)\xi}\, \rho^0\,\frac{A_x}{2} \left[ e^{i(kz+\theta)} - e^{-i(kz+\theta)}\right] \\
          {} & = & \ds - \frac{4}{2-3\nu+(2-\nu)\xi}\,\rho^0 A_x \sin(kz+\theta).
        \end{array}
      \end{equation}
      Combination of equations (\ref{eqn:InstMode}), (\ref{eqn:Psix}) and (\ref{eqn:Psiy}) yields the final expression of the instability mode:
      \begin{equation}\label{eqn:InstabMode}
        \left\{
          \begin{array}{lll}
            \phi_x(x,z) & = & A_xe^{\xi kx} \cos(kz+\theta) \\
            \phi_y(x,z) & = & A_ye^{\xi kx} \sin(kz+\theta)
          \end{array}
        \right.
        \quad {\rm where} \quad \frac{A_y}{A_x} \equiv - \frac{4}{2-3\nu+(2-\nu)\lambda/{k}}\,\rho^0.
      \end{equation}
      Equations (\ref{eqn:InstabMode}) show that {\it the instability mode corresponds to a perturbed crack front having the shape of an elliptic helix, of axis coinciding with the present unperturbed crack front and ``amplitudes'' $A_xe^{\xi kx}$, $A_ye^{\xi kx}$ in the directions $x$ and $y$ respectively, growing in proportion, and exponentially with the distance of propagation $x$}. This conclusion, and even the precise relation (\ref{eqn:InstabMode})$_3$ between the amplitudes $A_x$ and $A_y$, are exactly the same as in the case of a critical energy-release-rate independent of the mode mixity ratio, see \citep{LKL11}.\footnote{This is because the equations derived in this Subsection result only from the PLS which bears no relationship to the energy-release-rate.} Fig. \ref{fig:GeomDeform} illustrates the geometry of the instability mode.

\begin{figure}[h]
  \centering
     \subfigure[\label{fig:GeomSurf} Geometry of perturbed surface.]
       {
         \epsfig{figure=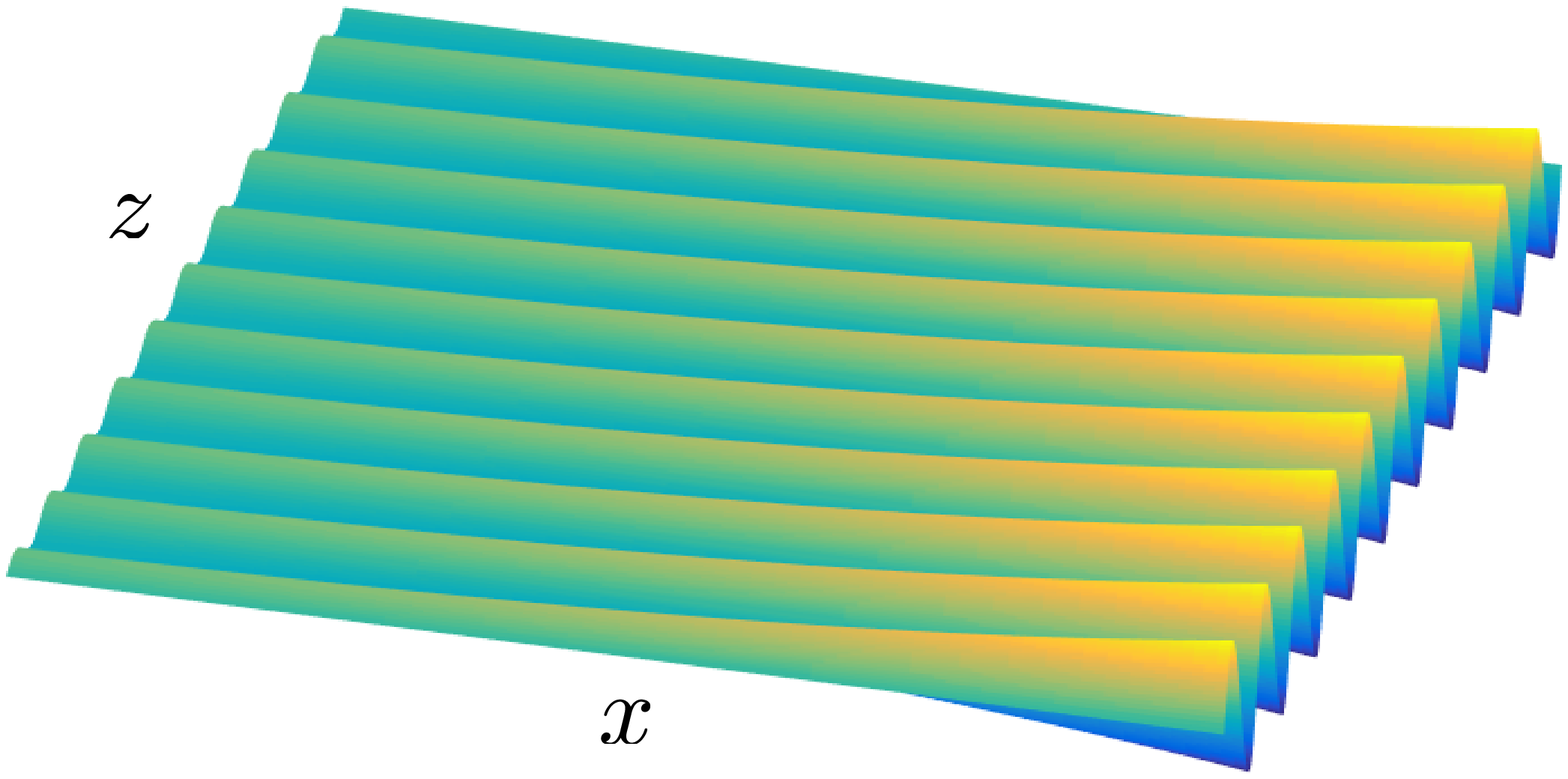,width=0.45\textwidth}
       }
     \subfigure[\label{fig:GeomFront} Geometry of perturbed front.]
       {
         \epsfig{figure=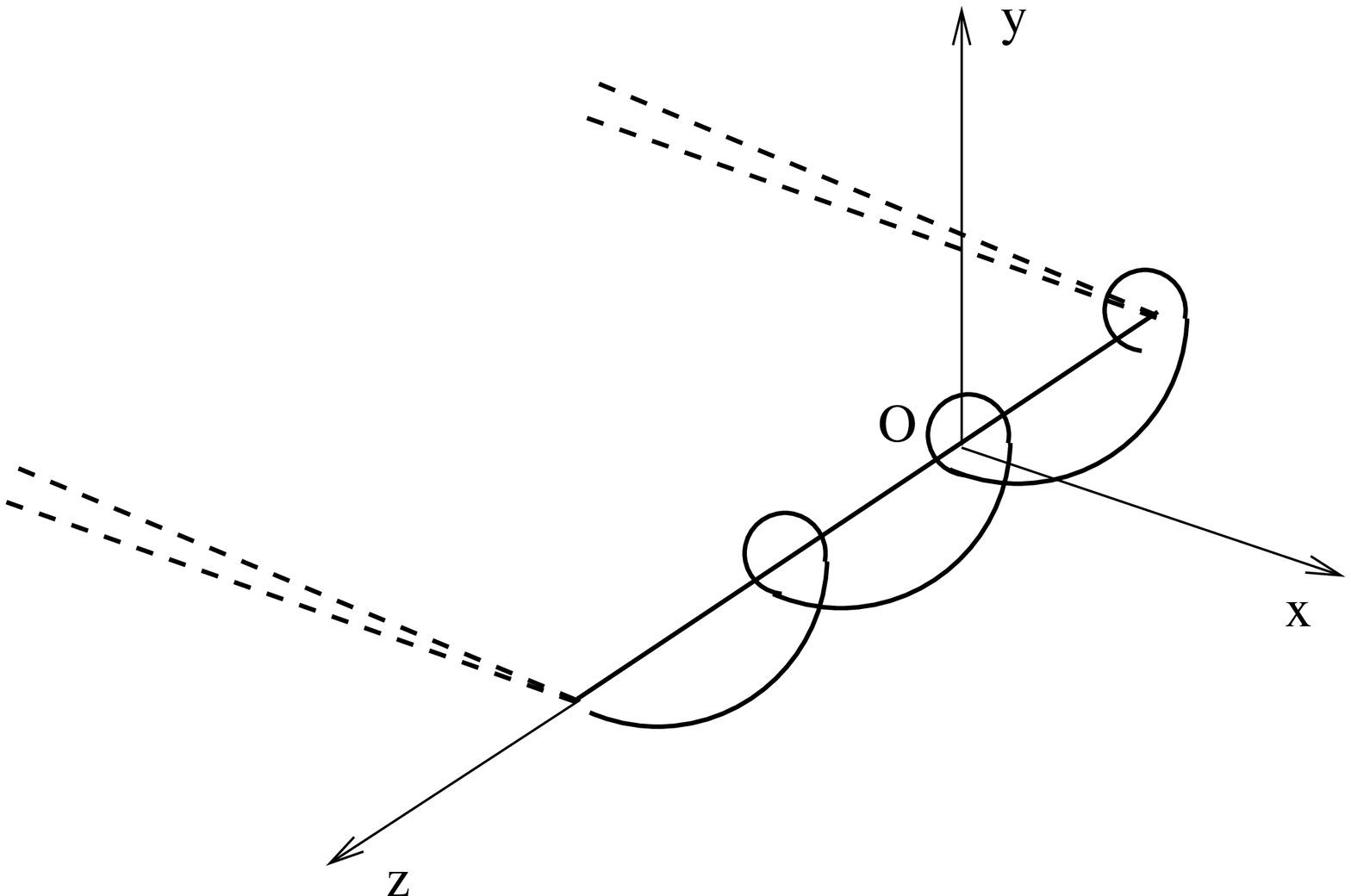,width=0.45\textwidth}
       }
     \caption{\label{fig:GeomDeform} Geometry of the surface and front of the crack in the perturbed configuration corresponding to an instability mode.}
\end{figure}

  \item {\it Case 2: $\rho^0 = \rho^{\rm cr}$}.\vskip2mm
   Then the solution of equation (\ref{eqn:EqOnXi}) is $\xi=\frac{\lambda}{|k|}=0$ so that $\lambda=0$ and $k$ is arbitrary. The out-of-plane perturbation $\phi_y$ is independent of the distance $x$ of propagation, and since the components of its Fourier transform in the direction $z$ of the crack front are totally unconstrained, it varies arbitrarily in this direction. This means that {\it coplanar propagation is (neutrally) unstable versus arbitrary out-of-plane deviations from coplanarity}. On the other hand, once the out-of-plane perturbation $\phi_y$ is fixed, the in-plane perturbation $\phi_x$ is also fixed and determined by equation (\ref{eqn:PhiyVsPhix}) (where $\xi=0$), which connects the two perturbations {\it via} the PLS.
\end{itemize}


\subsection{Numerical illustrations}\label{subsec:Mode13Num}

In this Section we provide numerical examples of predictions of the values of the critical mode mixity ratio and the normalized growth rate of unstable perturbations. A wide range of possible material parameters is considered, without any reference to a specific material.

\begin{figure}[h]
  \centering
     \subfigure[\label{fig:SmallGamma} Small values of $ \gamma$.]
       {
         \epsfig{figure=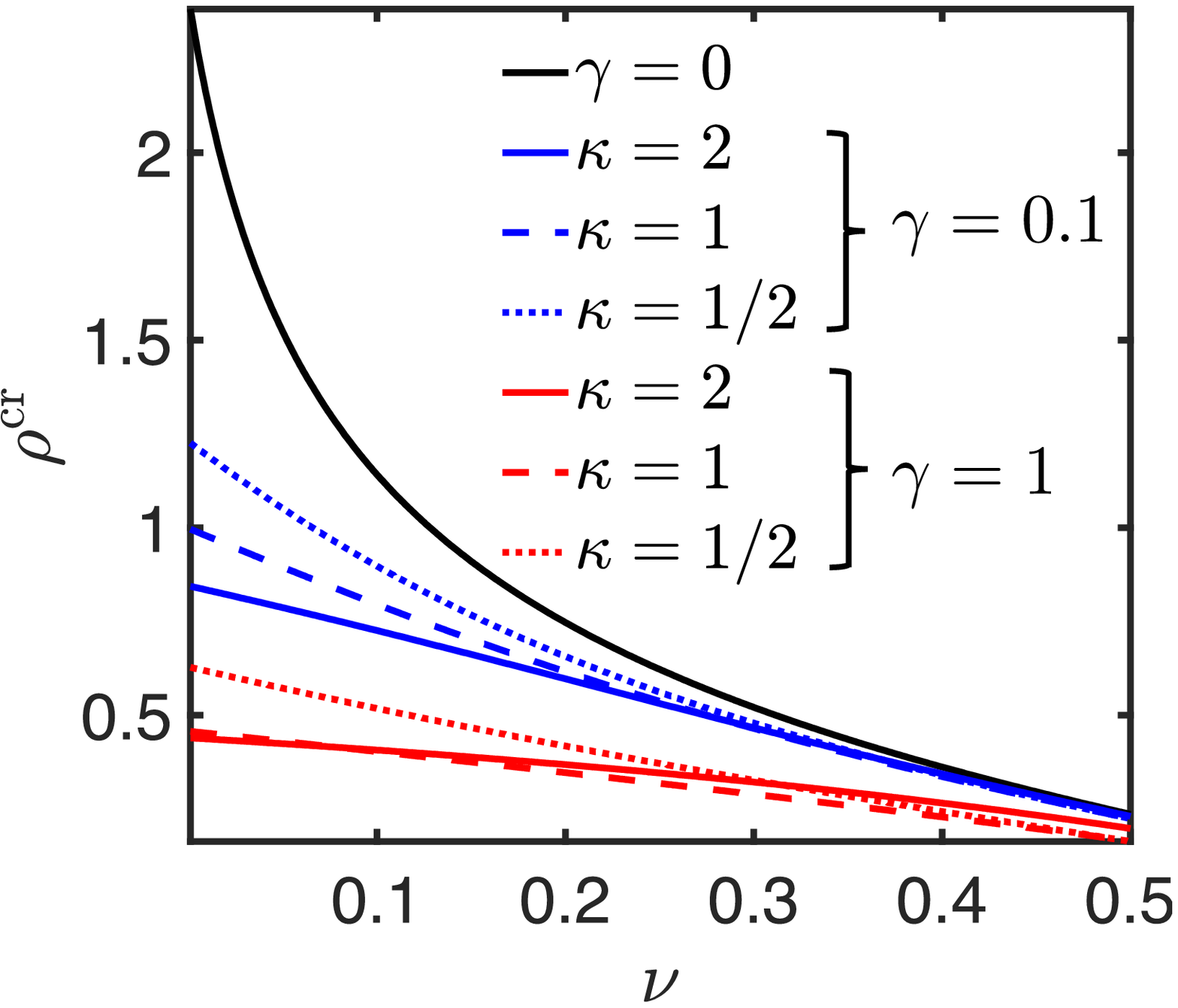,width=0.45\textwidth}
       }
     \subfigure[\label{fig:LargeGamma} Large values of $ \gamma$.]
       {
         \epsfig{figure=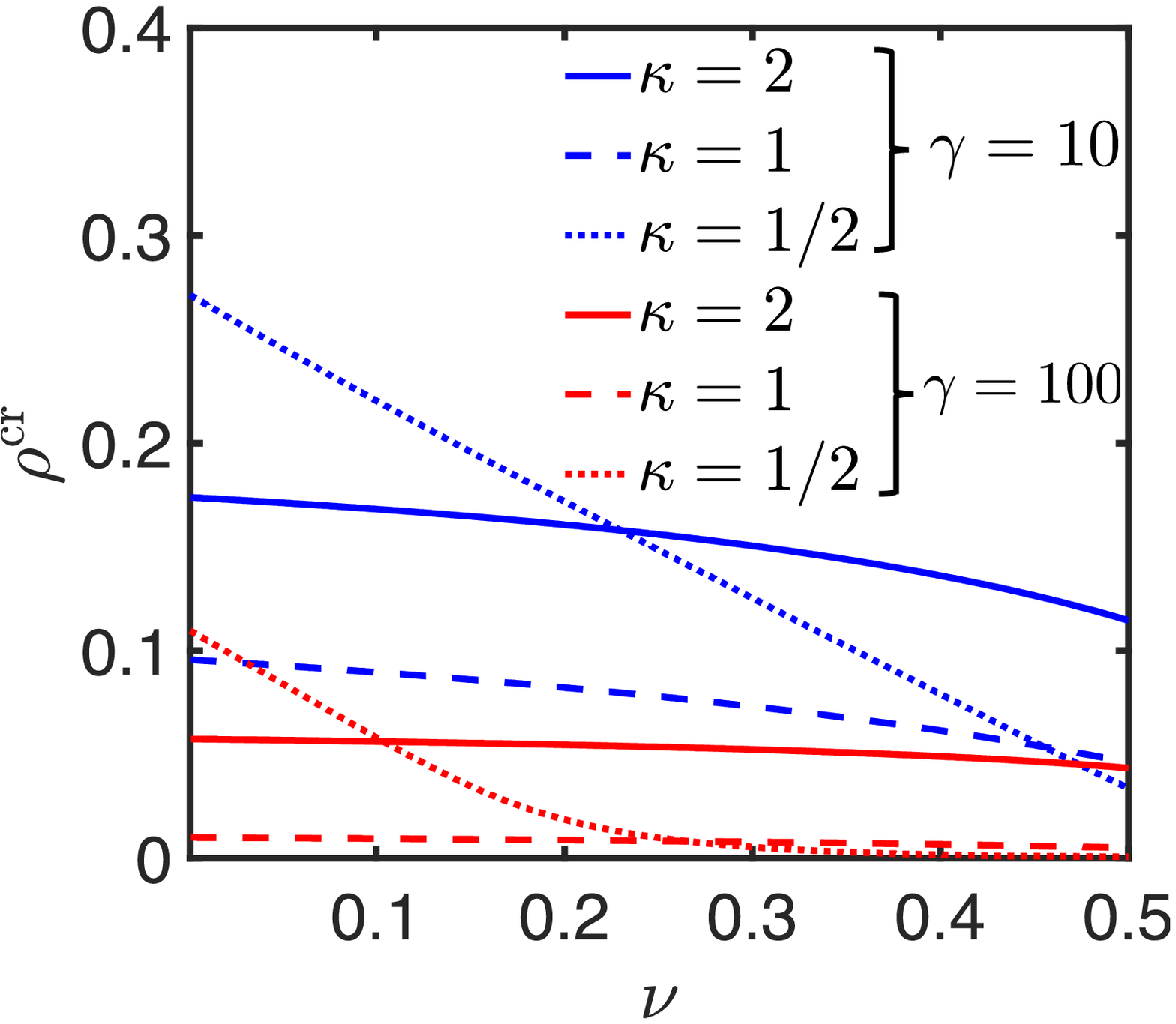,width=0.45\textwidth}
       }
     \caption{\label{fig:RhocVsNu} Critical mode mixity ratio $\rho^{\rm cr}$ versus Poisson's ratio $\nu$, for various values of the parameters $\gamma$ and $\kappa$.}
\end{figure}

Fig. \ref{fig:RhocVsNu} illustrates the dependence of the critical mode mixity ratio $\rho^{\rm cr}$ - obtained by numerically solving the equation $N(\rho^{\rm cr};0)=0$ - upon Poisson's ratio $\nu$, for several values of $\gamma$ and $\kappa$.

It clearly appears that the dependence of $G_c$ upon $\rho$ has a major impact upon the value of the critical mode mixity ratio: the larger the value of $\gamma$, the smaller that of $\rho^{\rm cr}$, whereas the effect of $\kappa$ is more complex and depends upon the value of $\nu$.

More precisely, Fig. \ref{fig:RhocVsNu} seems to show that $\rho^{\rm cr} \ll 1$ for large values of $\gamma$. Quite remarkably, an approximate analytical expression of this critical value may be found in this case. Indeed when $\rho \ll 1$, the expression (\ref{eqn:EqOnXi})$_2$ of $N(\rho;\xi)$ becomes for $\xi=0$, using the approximations $-(1-\nu)(2-3\nu)+[...]\rho^2 \simeq -(1-\nu)(2-3\nu)$, $1-\nu+\rho^2 \simeq 1-\nu$, $4-5\nu+2(...)\rho^2 \simeq 4-5\nu:$
\begin{equation*}
  N(\rho;0) \simeq -(1-\nu)(2-3\nu) + \frac{\kappa\gamma\rho^{\kappa}}{1+\gamma\rho^{\kappa}} (1-\nu)(4-5\nu).
\end{equation*}
The solution $\rho^{\rm cr}$ of the equation $N(\rho^0;0)=0$ is therefore given by
\begin{equation}\label{eqn:RhoCrLargeGamma}
  \rho^{\rm cr} \simeq \left[ \frac{2-3\nu}{(4-5\nu)\kappa-(2-3\nu)}\, \frac{1}{\gamma} \right]^{1/\kappa} \quad\quad (\gamma\gg 1).
\end{equation}
Expression (\ref{eqn:RhoCrLargeGamma}) applies provided that $(4-5\nu)\kappa-(2-3\nu)$ is positive, which is true for $\kappa \ge 1/2$ (with $0<\nu<1/2$). It confirms that $\rho^{\rm cr}$ goes to zero when $\gamma$ goes to infinity, as was anticipated from Fig. \ref{fig:RhocVsNu}. It also implies that when both $\gamma$ and $\kappa$ are large, $\rho^{\rm cr}$ depends only marginally on $\nu$ since the exponent $1/\kappa$ becomes small.

\begin{figure}[h]
\centerline{\psfig{figure=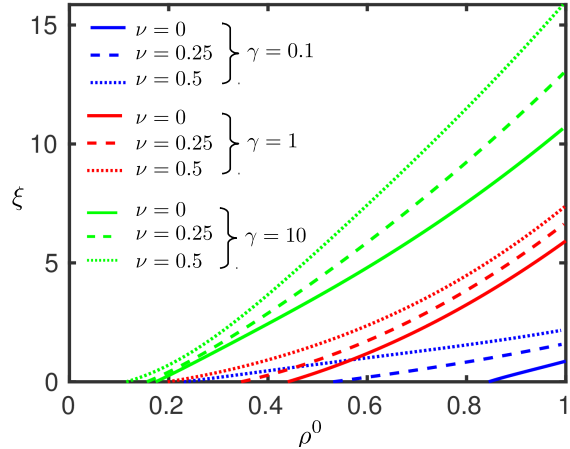,height=6cm}}
\caption{Normalized growth rate $\xi$ of the instability mode versus the unperturbed mode mixity ratio $\rho^0$, for $\kappa = 2$ and various values of $\nu$ and $\gamma$.}
\label{fig:XiVsRho}
\end{figure}

Fig. \ref{fig:XiVsRho} illustrates the dependence of the normalized growth rate $\xi$ of the instability mode - obtained by numerically solving equation (\ref{eqn:EqOnXi}) - upon the unperturbed mode mixity ratio $\rho^0$, for a few values of the material parameters. One clearly sees that $\xi$ takes positive values (implying an instability) only for values of $\rho^0$ larger than the material-dependent critical value $\rho^{\rm cr}$.

Finally, we make a tentative comparison with the experimental measurements of \cite{LMR10}. Assuming a value of $ \kappa = 2 $, the value of $ \gamma $ leading to a best fit (for small $\rho$) of the experimental crack initiation curve is $ \gamma \approx 25$ (see Fig. \ref{fig:KIIIbyKIcVsKIbyKIc}). The critical threshold of instability for these material parameters is $\rho^{\rm cr} \sim 0.1 $, which is in the range of loading considered in the experiments.


\section{Perturbations of the energy-release-rate and the mode mixity ratio for stationary perturbations}\label{sec:PertStat}

The case of stationary perturbations (independent of $x$) satisfying the PLS (equation (\ref{eqn:PhiyVsPhix})) is of special interest since they correspond to instability modes at the onset of the instability ($\rho^0=\rho^{\rm cr}$, $\lambda=0$). It so happens that simple and explicit expressions, in the physical space rather than that of Fourier, may be obtained in this special case for the perturbations $\delta G$ and $\delta \rho$ of the energy-release-rate and mode mixity ratio.

Perturbations independent of $x$ are of the form (\ref{eqn:InstMode}) with $\lambda\equiv 0$, $\psi_x\equiv\phi_x$, $\psi_y\equiv\phi_y$; hence provided they satisfy the PLS, equations (\ref{eqn:DeltaGDeltaRho}) are applicable with $\xi\equiv {\lambda}/{|k|}=0$ and yield
\begin{equation*}
  \left\{
    \begin{array}{lll}
      \ds \frac{\widehat{\delta G}(k)}{G^0} & = & - f(\rho^0;0)\,|k|\widehat{\phi_x}(k) \\
      \widehat{\delta\rho}(k) & = & - g(\rho^0;0)\,|k|\widehat{\phi_x}(k)
    \end{array}
  \right.
\end{equation*}
where
\begin{equation}\label{eqn:FGXiZero}
  \left\{
    \begin{array}{lll}
      f(\rho;0) & \equiv & \ds \frac{1-\nu}{1-\nu+\rho^2} \left[ 1 - \frac{3(2-\nu)-4\sqrt{2}(1-2\nu)}
                                    {(1-\nu)\left( 2-3\nu \right)}\,\rho^2 \right] \\[5mm]
      g(\rho;0) & \equiv & \ds \frac{\rho}{2-3\nu} \left[ 4-5\nu
                                    +2\left( 4-\sqrt{2}\,\frac{1-2\nu}{1-\nu} \right)\rho^2 \right].
    \end{array}
  \right.
\end{equation}
But equation (\ref{eqn:PhiyVsPhix}) expressing the PLS implies that
\begin{equation*}
  |k|\widehat{\phi_x}(k) = -i\, \frac{2-3\nu}{4\rho^0}\, k\widehat{\phi_y}(k);
\end{equation*}
hence $\widehat{\delta G}$ and $\widehat{\delta\rho}$ may be rewritten in the form
\begin{equation*}
  \left\{
    \begin{array}{lllll}
      \ds \frac{\widehat{\delta G}(k)}{G^0} & = & \ds \frac{2-3\nu}{4\rho^0}\, f(\rho^0;0)\,.\,ik\widehat{\phi_y}(k)
      & = & \ds \frac{2-3\nu}{4\rho^0}\, f(\rho^0;0)\widehat{\ \frac{d\phi_y}{dz}\ }(k)  \\[5mm]
      \widehat{\delta\rho}(k) & = & \ds \frac{2-3\nu}{4\rho^0}\, g(\rho^0;0)\,.\,ik\widehat{\phi_y}(k)
      & = & \ds \frac{2-3\nu}{4\rho^0}\, g(\rho^0;0)\widehat{\ \frac{d\phi_y}{dz}\ }(k).
    \end{array}
  \right.
\end{equation*}
The factors $\frac{2-3\nu}{4\rho^0}\, f(\rho^0;0)$ and $\frac{2-3\nu}{4\rho^0}\, g(\rho^0;0)$ here are independent of the wavenumber $k$. Therefore Fourier inversion of these relations yields directly
\begin{equation}\label{eqn:DGDRSpecCase}
  \left\{
    \begin{array}{lll}
      \ds \frac{\delta G(z)}{G^0} & = & \ds \frac{2-3\nu}{4\rho^0}\, f(\rho^0;0)\,\frac{d\phi_y}{dz}(z)  \\[5mm]
      \delta\rho(z) & = & \ds \frac{2-3\nu}{4\rho^0}\, g(\rho^0;0)\,\frac{d\phi_y}{dz}(z)\,.
    \end{array}
  \right.
\end{equation}

\begin{figure}[h]
\centerline{\psfig{figure=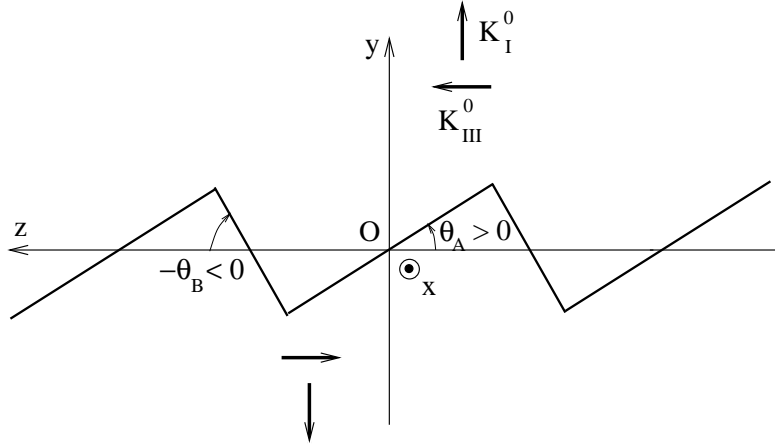,height=6cm}} \caption{``Factory-roof-shaped'' crack loaded in mode I+III.}
\label{fig:FactoryRoof}
\end{figure}

These remarkable equations directly relate the perturbations of the energy-release-rate and the mode mixity ratio to the local slope $d\phi_y/dz$ of the crack front in the plane perpendicular to the direction of propagation. They have interesting consequences. For instance, consider (Fig. \ref{fig:FactoryRoof}) the case of a crack loaded in mode I+III (with $K_{III}^0>0$), with a surface shaped like a ``factory roof'', that is consisting of ``facets of type A'' obtained through rotation of the original crack plane $Ozx$ by an angle $\theta_A>0$ about the direction $x$ of propagation, and ``facets of type B'' obtained through a similar rotation but with an angle $-\theta_B<0$.\footnote{The denominations ``facets of type A'', ``facets of type B'' are due to \cite{HP79}.} Assume that these facets have reached a stationary state and satisfy the PLS. Then equations (\ref{eqn:DGDRSpecCase}) apply, and imply that {\it the perturbations of the energy-release-rate and the mode mixity ratio} (and hence the energy-release-rate and the mode mixity ratio themselves) {\it take constant values over facets of each type}, given with obvious notations by
\begin{equation}\label{eqn:DGDRFactoryRoof}
  \left\{
    \begin{array}{lll}
      \ds \frac{\delta G_A}{G^0} & = & \ds - \frac{2-3\nu}{4\rho^0}\, f(\rho^0;0)\,\theta_A \\[5mm]
      \delta\rho_A & = & \ds - \frac{2-3\nu}{4\rho^0}\, g(\rho^0;0)\,\theta_A
    \end{array}
  \right.
  \quad ; \quad
  \left\{
    \begin{array}{lll}
      \ds \frac{\delta G_B}{G^0} & = & \ds \frac{2-3\nu}{4\rho^0}\, f(\rho^0;0)\,\theta_B \\[5mm]
      \delta\rho_B & = & \ds \frac{2-3\nu}{4\rho^0}\, g(\rho^0;0)\,\theta_B.
    \end{array}
  \right.
\end{equation}

The case of predominant mode I is of special interest. In such a case the mode mixity ratio is much smaller than unity so that the terms proportional to its square in the expressions (\ref{eqn:FGXiZero}) of $f(\rho;0)$ and $g(\rho;0)$ may be disregarded, and the expressions (\ref{eqn:DGDRSpecCase}) of $\delta G$ and $\delta \rho$ take the following wonderfully simple form, for arbitrary out-of-plane perturbations $\phi_y$:
\begin{equation}\label{eqn:DGDRSpecSpecCase}
  \left\{
    \begin{array}{lll}
      \ds \frac{\delta G(z)}{G^0} & = & \ds \frac{2-3\nu}{4\rho^0}\,\frac{d\phi_y}{dz}(z)  \\[5mm]
      \delta\rho(z) & = & \ds \frac{1}{4}\,(4-5\nu)\,\frac{d\phi_y}{dz}(z).
    \end{array}
  \right.
\end{equation}

A simple and interesting application of formula (\ref{eqn:DGDRSpecSpecCase}$_2$) consists in approximately predicting the inclination angle $\theta_A$ of facets of type A on a factory-roof-shaped crack surface (Fig.~\ref{fig:FactoryRoof}) having reached its stationary state. If we adopt, following \cite{CP96}, the assumption that facets of this type are shear-free, the resulting condition $\rho_{A} = \rho^0 + \delta\rho_{A} = 0$, combined with equation~\eqref{eqn:DGDRSpecSpecCase}$_2$, yields
\begin{equation}\label{eqn:PCmodified}
\theta_{A} = \frac{\rho^0}{1 - 5 \nu/4}\,.
\end{equation}
This expression calls for a few comments.
\begin{itemize}
  \item Since \cite{MGW98}'s formulae used in the derivation of equation (\ref{eqn:PCmodified}) are accurate only to first order in the perturbations of the crack surface and front, the prediction of this equation is applicable only for small inclinations of facets of type A, resulting from small mode mixity ratios $\rho^0$.
  \item The inclination angle $\theta_A$ given by equation (\ref{eqn:PCmodified}) differs from that predicted by \cite{CP96}'s well-known formula, $\theta_{A}^{\rm CP} = \rho^0/(1-2 \nu)$ (for small $\rho^0$), due to different hypotheses and methods of calculation. \cite{CP96}'s formula was for incipient facets instead of stationary ones, and was based on an elementary calculation involving only the asymptotic stress field prior to any segmentation, rather than evaluation of the distribution of the local mode III SIF along an already fragmented crack front.
  \item Equation (\ref{eqn:PCmodified}) provides a lower estimate of the inclination of facets of type A than \cite{CP96}'s formula, agreeing better with the results of both experiments and numerical computations (based on a phase-field model) performed by \cite{CCLNPK15}.
  \item Formula (\ref{eqn:PCmodified}) eliminates the physically groundless divergence of the inclination angle predicted by \cite{CP96}'s formula in the limit of incompressible materials ($\nu \rightarrow 1/2$).
\end{itemize}

A final remark pertains to the practical relevance of the factory roof geometry. This geometry has been observed in many experiments. It represents an asymptotically exact stationary solution of the linearized equations of crack propagation for $\rho^0=\rho^{\rm cr}$; but, as remarked in Subsection \ref{subsec:Mode13GeomMode}, the same is in fact true of {\it all} smooth small-amplitude out-of-plane perturbations.\footnote{This basically stems from the fact that all Fourier modes bifurcate at the same critical value $\rho^{\rm cr}$ for which the instability growth rate vanishes (that is, the bifurcation is infinite dimensional). Hence for $\rho^0=\rho^{\rm cr}$ any out-of-plane perturbation, being a sum of Fourier modes of different amplitudes, is a stationary solution.} Understanding why the factory roof shape is dynamically selected in experiments and phase-field simulations, rather than any other shape compatible with linear stability analysis, is beyond the scope of this work. It would most likely require to consider both nonlinear effects and the role of the process zone scale, which has been proposed by \cite{PK10} to modify the PLS by making the local mode II SIF dependent on out-of-plane curvature of the crack front, in a way analogous to surface tension in other interfacial pattern forming instabilities.


\section{Summary of results and perspectives}\label{sec:Concl}

In this paper, we presented an extension of our previous linear stability analysis of coplanar crack propagation in mixed-mode I+III \citep{LKL11}, including a heuristic hypothesis of dependence of the critical energy-release-rate $G_c$ upon the ratio $\rho \equiv K_{III}^0/K_I^0$ of the unperturbed mode III to mode I SIFs. The essential aim was to see whether accounting for a shear dependence of the fracture energy could lead to a significant decrease of the theoretical threshold $\rho^{\rm cr}$ above which coplanar propagation becomes unstable, and generate values low enough to become compatible with those generally observed experimentally.

Section \ref{sec:PertCrack} was first devoted to general hypotheses and notations and some fundamental formulae used throughout the paper, providing the SIFs along the front of a slightly perturbed semi-infinite crack in an infinite body. These formulae were taken from the work of \cite{GR86} for the in-plane perturbation of the crack front, and the work of \cite{MGW98} (duly completed by that of \cite{LKL11}) for the out-of-plane perturbation of the crack surface.

Section \ref{sec:Mode13StabAnal} then presented the extended stability analysis. Apart from the additional hypothesis of dependence of $G_c$ upon $\rho$, the new analysis relied on the same basic elements as our previous one \citep{LKL11}; that is, in addition to the technical elements explained in the preceding Section, use of a double propagation criterion combining \cite{G20}'s energetic condition and \cite{GS74}'s principle of local symmetry, enforced at all instants and all points along the perturbed crack front.
The geometrical features of instability modes above the threshold were found to be the same as for a constant $G_c$, the perturbed crack front still assuming the shape of an elliptic helix with size growing exponentially with the distance of propagation.
More importantly, the dependence of $G_c$ upon $\rho$ was found to generate a significant decrease of the theoretical threshold $\rho^{\rm cr}$, as was hoped. Furthermore, an estimate of this threshold obtained by assuming a simple functional form of $G_c(\rho)$ with parameters fitted to experimentally measured crack initiation curves of \cite{LMR10} (cf. Fig. \ref{fig:KIIIbyKIcVsKIbyKIc}) was found to be compatible with the range of $\rho$-values for which fragmentation was observed in these experiments.

Section \ref{sec:PertStat} was devoted to a complementary study of some remarkable properties of the SIFs and energy-release-rate along the perturbed crack front, in the special case of stationary perturbations satisfying the principle of local symmetry. Such perturbations are interesting in that they correspond to instability modes in near-threshold conditions, with negligible rate of growth. The perturbations of the SIFs and the energy-release-rate were shown to be both proportional to the local slope of the perturbed front in the plane perpendicular to the direction of propagation. For instance, for a crack surface having the shape of a ``factory roof'', the SIFs of mode I and III and the energy-release-rate are all constant on every flat portion of the ``roof''. It is hoped that the formulae derived will be useful in the future to analyze and understand the inclinations of the facets of type A and B observed experimentally, as functions of the mode mixity ratio and the material properties.

This work opens several perspectives:
\begin{itemize}
  \item Theoretical predictions should be compared with the results of more fracture experiments performed under mixed mode I+III conditions. The first aim of experiments would be to study the dependence of the fracture energy upon the mode mixity ratio, and if possible precisely quantify this dependence independently of any observation of the threshold value of this ratio. One could then deduce the theoretical value of the threshold from this observed dependence, and compare it to the experimental value.
  \item The theoretical value of the threshold should be compared to the results of phase-field simulations analogous to those of \cite{PK10} and \cite{CCLNPK15}, but including an extra dependence of the fracture energy upon the mode mixity ratio. The aim of such simulations would be to account for nonlinear geometrical effects and the role of the process zone scale disregarded in the theoretical linear stability analysis, but which might be of primary importance, especially for disconnected facets.
  \item It would be interesting to investigate the question of the physical mechanisms responsible for the dependence of the fracture energy upon the mode mixity ratio, left completely open by the present work where this dependence was simply postulated as a heuristic hypothesis.
  \item Finally the stability analysis should be extended to general mixed-mode I+II+III conditions, since experiments almost always include some small, uncontrolled mode II loading component. The extension will inevitably be limited to small values of the mode II SIF applied remotely, since large values would result in a large general kink of the crack prohibiting the use of \cite{MGW98}'s geometrically linearized formulae.
\end{itemize}


{\bf Acknowledgements}

The authors wish to express their sincere thanks to Professor K. Ravi-Chandar of the University of Austin, Texas for several very stimulating discussions. A.K. acknowledges support of Grant DEFG02-07ER46400 from the US Department of Energy, Office of Basic Energy Sciences. L.P. and A.V. acknowledge support of the City of Paris through the Emergence Program.


\include{AppInstabGcVarPart1}

\end{document}

%% file: AppInstabGcVarPart1.tex
\appendix

\section{Appendix : calculation of the Fourier transform of $\delta_y K_I^{\rm skew}$}\label{app:FourierDeltaK1Skew}

To calculate the Fourier transform $\widehat{\delta_y K_I^{\rm skew}}(x,k)$ for an out-of-plane perturbation of the crack $\phi_y(x,z)$ of the form (\ref{eqn:InstMode})$_2$, the first task is to evaluate $\delta_y K_I^{\rm skew}(x,z)$ for the perturbation $e^{\lambda x}\,\widehat{\psi_y}(k)e^{ikz}$. For this perturbation equation (\ref{eqn:dyK1skew}) yields
\begin{equation*}
  \begin{array}{l}
    [\delta_y K_I^{\rm skew}(x,z)]_{|\,\phi_y=e^{\lambda x}\,\widehat{\psi_y}(k)e^{ikz}} \\[3mm]
    \ds \quad \quad = \frac{\sqrt{2}}{4\pi} \, \frac{1-2\nu}{1-\nu} \, K_{III}^0 {\rm Re} \left\{ik \widehat{\psi_y}(k) \int_{-\infty}^{x} dx' \int_{-\infty}^{+\infty} \frac{e^{\lambda x'}e^{ikz'}}{(x-x')^{1/2}\left[x-x'+i(z-z')\right]^{3/2}} \, dz' \right\} \\[3mm]
    \ds \quad \quad = \frac{\sqrt{2}}{4\pi} \, \frac{1-2\nu}{1-\nu} \, K_{III}^0 k\,e^{\lambda x}\,{\rm Re} \left\{i \widehat{\psi_y}(k)
                      \int_{0}^{+\infty} \frac{e^{-\lambda v}}{\sqrt{v}} \,dv
                      \int_{-\infty}^{+\infty} \frac{e^{ikz'}}{\left[v+i(z-z')\right]^{3/2}} \, dz' \right\}
  \end{array}
\end{equation*}
where the change of variable $v\equiv x-x'$ has been used. The integral over $z'$ here has been evaluated by \cite{LKL11} using complex analysis, and the result is:
\begin{equation*}
  \int_{-\infty}^{+\infty} \frac{e^{ikz'}}{\left[v+i(z-z')\right]^{3/2}} \, dz' = 4H(-k)\sqrt{\pi |k|}\,e^{-|k|v}e^{ikz}
\end{equation*}
where $H$ is Heaviside's function; it follows that
\begin{equation*}
  \begin{array}{l}
    [\delta_y K_I^{\rm skew}(x,z)]_{|\,\phi_y=e^{\lambda x}\,\widehat{\psi_y}(k)e^{ikz}} \\[3mm]
    \ds \quad \quad = \sqrt{\frac{2}{\pi}} \, \frac{1-2\nu}{1-\nu} \, K_{III}^0 k\sqrt{|k|}\,H(-k)\,e^{\lambda x} \,{\rm Re} \left[i \widehat{\psi_y}(k)e^{ikz} \int_{0}^{+\infty} \frac{e^{-(|k|+\lambda)v}}{\sqrt{v}} \,dv \right].
  \end{array}
\end{equation*}
Now use of the change of variable $w \equiv \sqrt{(|k|+\lambda)v}$ yields
\begin{equation*}
  \int_{0}^{+\infty} \frac{e^{-(|k|+\lambda)v}}{\sqrt{v}} \,dv = \frac{2}{\sqrt{|k|+\lambda}} \int_0^{+\infty} e^{-w^2}dw = \sqrt{\frac{\pi}{|k|+\lambda}}
                                                           = \sqrt{\frac{\pi}{|k|}} \, F(\xi)
\end{equation*}
where $\xi$ and $F(\xi)$ are defined by equations (\ref{eqn:DefXi}) and (\ref{eqn:FourierDeltaK1Skew})$_2$ of the text. Therefore
\begin{equation*}
    [\delta_y K_I^{\rm skew}(x,z)]_{|\,\phi_y=e^{\lambda x}\widehat{\psi_y}(k)e^{ikz}} =
    \sqrt{2}\, \frac{1-2\nu}{1-\nu} \, K_{III}^0 kH(-k)F(\xi)\,e^{\lambda x} \,{\rm Re} \left[i \widehat{\psi_y}(k)e^{ikz} \right].
\end{equation*}

For the value $-k$, one has $\widehat{\psi_y}(-k) = \overline{\widehat{\psi_y}(k)}$ since $\psi_y(z)$ is real, so that
\begin{equation*}
  \begin{array}{lll}
    [\delta_y K_I^{\rm skew}(x,z)]_{|\,\phi_y=e^{\lambda x}\widehat{\psi_y}(-k)e^{-ikz}} & = &
        \ds \sqrt{2}\, \frac{1-2\nu}{1-\nu} \, K_{III}^0 (-k)H(k)F(\xi)\,e^{\lambda x} \,{\rm Re} \left[i \overline{\widehat{\psi_y}(k)}\,)e^{-ikz} \right] \\[3mm]
    {} & = & \ds \sqrt{2}\, \frac{1-2\nu}{1-\nu} \, K_{III}^0 kH(k)F(\xi)\,e^{\lambda x} \,{\rm Re} \left[i \widehat{\psi_y}(k)e^{ikz} \right].
  \end{array}
\end{equation*}
For the perturbation $e^{\lambda x}.\,\frac{1}{2}\left[\widehat{\psi_y}(k)e^{ikz}+\widehat{\psi_y}(-k)e^{-ikz}\right]$, it then follows from use of the relation $H(k)+H(-k)=1$ that
\begin{equation*}
  \begin{array}{l}
     [\delta_y K_I^{\rm skew}(x,z)]_{|\,\phi_y=e^{\lambda x}.\,\frac{1}{2}\left[\widehat{\psi_y}(k)e^{ikz}+\widehat{\psi_y}(-k)e^{-ikz}\right]} \\[3mm]
     \quad \quad \ds = \frac{1}{\sqrt{2}}\, \frac{1-2\nu}{1-\nu} \, K_{III}^0 kF(\xi)\,e^{\lambda x} \,{\rm Re} \left[i \widehat{\psi_y}(k)e^{ikz} \right] \\[3mm]
     \quad \quad \ds = \frac{1}{\sqrt{2}}\, \frac{1-2\nu}{1-\nu} \, K_{III}^0 \,e^{\lambda x}F(\xi).\frac{1}{2} \left[ik \widehat{\psi_y}(k)e^{ikz}
                       - ik \widehat{\psi_y}(-k)e^{-ikz} \right]
   \end{array}
\end{equation*}
where the property $\widehat{\psi_y}(-k) = \overline{\widehat{\psi_y}(k)}$ has been used again.

Since the real perturbation defined by equation (\ref{eqn:InstMode})$_2$ may be written as
\begin{equation*}
  \phi_y(x,z) = \int_{-\infty}^{+\infty} e^{\lambda x}\,\widehat{\psi_y}(k)e^{ikz}dk
  = \int_{-\infty}^{+\infty} e^{\lambda x}.\,\frac{1}{2}\left[\widehat{\psi_y}(k)e^{ikz}+\widehat{\psi_y}(-k)e^{-ikz}\right] dk,
\end{equation*}
it follows that for this real perturbation
\begin{equation*}
  \begin{array}{lll}
     \delta_y K_I^{\rm skew}(x,z) & = & \ds \int_{-\infty}^{+\infty} \frac{1}{\sqrt{2}}\, \frac{1-2\nu}{1-\nu} \, K_{III}^0 \,e^{\lambda x}F(\xi).\frac{1}{2} \left[ik \widehat{\psi_y}(k)e^{ikz} - ik \widehat{\psi_y}(-k)e^{-ikz} \right] dk \\[3mm]
     {} & = & \ds \int_{-\infty}^{+\infty} \frac{1}{\sqrt{2}}\, \frac{1-2\nu}{1-\nu} \, K_{III}^0 \,e^{\lambda x}F(\xi).ik \widehat{\psi_y}(k)e^{ikz} dk.
   \end{array}
\end{equation*}
Comparison with the definition (\ref{eqn:DefFourier})$_1$ of the Fourier transform then shows that
\begin{equation*}
  \widehat{\delta_y K_I^{\rm skew}}(x,k) = \frac{i}{\sqrt{2}}\, \frac{1-2\nu}{1-\nu} \, K_{III}^0 \,e^{\lambda x}F(\xi) k \widehat{\psi_y}(k),
\end{equation*}
which is equation (\ref{eqn:FourierDeltaK1Skew}) of the text.

\section{Appendix : qualitative mathematical analysis of equation (\ref{eqn:EqOnXi}) on $\xi$}\label{app:MathStudy}

It is assumed in this Appendix that $0<\nu<1/2$ - these inequalities being always satisfied in practice although thermodynamic stability requirements warrant only that $-1 < \nu < 1/2$.

Equation (\ref{eqn:EqOnXi}) on $\xi$ is written in the form
\begin{equation*}
  \Phi(\rho^0;\xi) = 0 \quad {\rm where} \quad \Phi(\rho;\xi) \equiv \xi - \frac{N(\rho;\xi)}{D(\rho)}\,.
\end{equation*}

\subsection{Study of the function $N(\rho;0)$}\label{subapp:FunctionN0}

The function $N(\rho;0)$ is given by equation (\ref{eqn:EqOnXi})$_2$ of the text with $\xi=0$, that is:
\begin{equation*}
  \begin{array}{lll}
  N(\rho;0) & = & \ds -(1-\nu)(2-3\nu) + \left[ 3(2-\nu) - 4\sqrt{2}(1-2\nu) \right]\rho^2 \\[3mm]
      {} & {} & \ds + X(\rho) \left( 1-\nu+\rho^2 \right) \left[ 4-5\nu+2\left( 4-\sqrt{2}\frac{1-2\nu}{1-\nu} \right)\rho^2 \right]
  \end{array}
\end{equation*}
where $X(\rho)$ is given by equation (\ref{eqn:EqOnXi})$_4$. The following properties follow:
\begin{itemize}
  \item At the point $\rho=0$, $N(\rho;0) = -(1-\nu)(2-3\nu)$ is negative.
  \item In the limit $\rho\rightarrow +\infty$, $X(\rho)$ becomes constant so that $N(\rho;0)$ is positively proportional to $\rho^4$ and therefore goes to $+\infty$.
  \item The quantities
  \begin{equation*}
    \left[ 3(2-\nu) - 4\sqrt{2}(1-2\nu) \right]\rho^2,\ X(\rho),\ 1-\nu+\rho^2,\ 4-5\nu+2\left( 4-\sqrt{2}\frac{1-2\nu}{1-\nu} \right)\rho^2
  \end{equation*}
   are all positive and increase with $\rho$, so that $N(\rho;0)$ is an increasing function of $\rho$.
\end{itemize}

\subsection{Consequences for the equation $N(\rho^0;0)=0$}\label{subapp:EqN0}

From the first two properties established in Subsection \ref{subapp:FunctionN0}, combined with the continuity of the function $N(\rho;0)$, one deduces that {\it the equation} $N(\rho^0;0)=0$ {\it admits at least one solution} $\rho^{\rm cr}$ {\it in the interval} $(0,+\infty)$. From the third, one concludes that {\it this solution is unique}, and that $N(\rho;0)<0$ for $\rho<\rho^{\rm cr}$, $N(\rho;0)>0$ for $\rho>\rho^{\rm cr}$.

\subsection{Positiveness of the denominator $D(\rho)$}\label{subapp:PosDenom}

The denominator $D(\rho)$ is given by equation (\ref{eqn:EqOnXi})$_3$ of the text. It follows from this expression, that of the quantity $X(\rho)$, equation (\ref{eqn:EqOnXi})$_4$, and the obvious inequality $\frac{\gamma\rho^{\kappa}}{1+\gamma\rho^{\kappa}} < 1$, that
\begin{equation*}
  \begin{array}{lll}
    D(\rho) & > & (1-\nu)(2-\nu) + (2+\nu)\rho^2 - \kappa\,\nu \left( 1-\nu+\rho^2 \right) \\
    {} & = & (1-\nu)(2-\nu-\kappa\nu) + (2+\nu-\kappa\nu)\rho^2.
  \end{array}
\end{equation*}
For the last quantity to be positive for all values of $\rho$, it suffices that $2-\nu-\kappa\nu$ be positive ($2+\nu-\kappa\nu$ is then also automatically positive). This is equivalent to $\nu < \frac{2}{1+\kappa}$. Since $\nu < \frac{1}{2}$, it suffices that $\frac{1}{2} \leq \frac{2}{1+\kappa}$ or equivalently $\kappa \leq 3$ for this to be true. {\it Such a condition, which seems experimentally reasonable \citep{LMR10}, will be assumed to be fulfilled in the sequel}. Then $D(\rho)$ is positive for all values of $\rho$.

\subsection{Study of the function $\Phi(\rho;\xi)$ for fixed $\rho$}\label{subapp:StudyPhi}

The following properties follow from the definition of the function $\Phi(\rho;\xi)$, the expressions (\ref{eqn:EqOnXi})$_{2,3}$ of $N(\rho;\xi)$ and $D(\rho)$, and the properties established in Subsections \ref{subapp:EqN0} and \ref{subapp:PosDenom}:
\begin{itemize}
  \item At the point $\xi=0$, $\Phi(\rho;0) = - {N(\rho;0)}/{D(\rho)}$ is positive if $\rho<\rho^{\rm cr}$, and negative if $\rho>\rho^{\rm cr}$.
  \item In the limit $\xi\rightarrow +\infty$, $\Phi(\rho;\xi)\sim\xi$ so that this function goes to $+\infty$.
  \item The derivative $\partial\Phi/\partial\xi$ is given by
  \begin{equation*}
    \frac{\partial\Phi}{\partial\xi}(\rho;\xi) = 1 - \frac{1}{D(\rho)} \left[ 2\sqrt{2}(1-2\nu)\rho^2
                                                    + X(\rho) \left( 1-\nu+\rho^2 \right) \sqrt{2}\,\frac{1-2\nu}{1-\nu}\,\rho^2 \right](1+\xi)^{-3/2};
  \end{equation*}
  since the quantities $D(\rho)$ and $[...]$ are positive, this derivative increases with $\xi$.
\end{itemize}

\subsection{Consequences for the equation $\Phi(\rho^0;\xi)=0$ for $\rho^0>\rho^{\rm cr}$}\label{subapp:EqOnXiLargeRho}

In the case where $\rho^0>\rho^{\rm cr}$, $\Phi(\rho^0;\xi)$ is negative at the point $\xi=0$ and goes to $+\infty$ in the limit $\xi\rightarrow +\infty$; hence {\it the equation} $\Phi(\rho^0;\xi)=0$ {\it on} $\xi$ {\it admits at least one solution in the interval} $(0,+\infty)$.

Let $\xi^1$ denote the first (smallest) solution. In the interval $(0,\xi^1)$, $\Phi(\rho^0;\xi)<0$; hence necessarily $\frac{\partial\Phi}{\partial\xi}(\rho^0;\xi^1) \geq 0$. Since $\frac{\partial\Phi}{\partial\xi}(\rho^0;\xi)$ is an increasing function of $\xi$, this implies that $\frac{\partial\Phi}{\partial\xi}(\rho^0;\xi) > 0$ for $\xi>\xi^1$. Hence $\Phi(\rho^0;\xi)$ increases with $\xi$ over the interval $(\xi^1,+\infty)$ and cannot vanish there. Thus {\it the solution of the equation} $\Phi(\rho^0;\xi)=0$ {\it on $\xi$ in the interval} $(0,+\infty)$ {\it is unique}.

\subsection{Positivity of the derivative $\frac{\partial\Phi}{\partial\xi}(\rho;0)$ for $\rho<\rho^{\rm cr}$}\label{subapp:PosDeriv}

We now assume that $\rho<\rho^{\rm cr}$ and look for conditions ensuring the positiveness of the derivative $\frac{\partial\Phi}{\partial\xi}(\rho;0)$. In order to avoid any ambiguity, we shall add to all desired, but not necessarily fulfilled inequalities, an interrogation mark `` ? '' after the inequality sign.

The above expression of $\frac{\partial\Phi}{\partial\xi}(\rho;\xi)$ shows that the inequality $\frac{\partial\Phi}{\partial\xi}(\rho;0) >? \ 0$ is equivalent to
\begin{equation*}
  2\sqrt{2}\,(1-2\nu)\rho^2 + X(\rho) \left( 1-\nu+\rho^2 \right) \sqrt{2}\,\frac{1-2\nu}{1-\nu}\,\rho^2
  <? \ D(\rho),
\end{equation*}
that is, account being taken of the expression (\ref{eqn:EqOnXi})$_{3}$ of $D(\rho)$:
\begin{equation*}
  (1-\nu)(2-\nu) + \left[ 2+\nu-2\sqrt{2}\,(1-2\nu) \right]\rho^2 >? \ X(\rho) \left( 1-\nu+\rho^2 \right) \left( \nu+\sqrt{2}\,\frac{1-2\nu}{1-\nu}\,\rho^2 \right).
\end{equation*}

This inequality cannot be true for all values of $\rho$ since in the limit $\rho \rightarrow +\infty$, the right-hand side goes to $+\infty$ proportionally to $\rho^4$ ($X(\rho)$ goes to a constant), more quickly than the left-hand side which is asymptotically proportional to $\rho^2$. But we only wish it to be true for values of $\rho$ smaller than $\rho^{\rm cr}$. For such values we have seen that $N(\rho;0)<0$, which implies by the above expression of $N(\rho;0)$ that
\begin{equation*}
  X(\rho) \left( 1-\nu+\rho^2 \right) <
  \frac{ (1-\nu)(2-3\nu) - \left[ 3(2-\nu) - 4\sqrt{2}(1-2\nu) \right]\rho^2 }{ 4-5\nu+2\left( 4-\sqrt{2}\,\frac{1-2\nu}{1-\nu} \right)\rho^2 }\,.
\end{equation*}
Hence to fulfill the desired inequality for values of $\rho$ smaller than $\rho^{\rm cr}$, it suffices to fulfill the inequality obtained by replacing the quantity $X(\rho) \left( 1-\nu+\rho^2 \right)$ by the upper bound just derived, that is:
\begin{equation*}
  \begin{array}{l}
    (1-\nu)(2-\nu) + \left[ 2+\nu-2\sqrt{2}\,(1-2\nu) \right]\rho^2 \\[3mm]
    \ds \quad\quad\quad\quad >? \ \frac{ (1-\nu)(2-3\nu) - \left[ 3(2-\nu) - 4\sqrt{2}(1-2\nu) \right]\rho^2 }{ 4-5\nu+2\left( 4-\sqrt{2}\,\frac{1-2\nu}{1-\nu} \right)\rho^2 } \left( \nu+\sqrt{2}\,\frac{1-2\nu}{1-\nu}\,\rho^2 \right).
  \end{array}
\end{equation*}

To study this inequality, rewrite it in the form
\begin{equation*}
  \begin{array}{l}
    (1-\nu)(2-\nu) + \left[ 2+\nu-2\sqrt{2}\,(1-2\nu) \right]\rho^2 \\[3mm]
    \ds \quad\quad\quad\quad >? \ \left\{ (1-\nu)(2-3\nu) - \left[ 3(2-\nu) - 4\sqrt{2}(1-2\nu) \right]\rho^2 \right\} {\mathcal H}(\rho)
  \end{array}
\end{equation*}
where ${\mathcal H}(\rho)$ is the homographic function of $\rho^2$ defined by
\begin{equation*}
  {\mathcal H}(\rho) \equiv \frac{ \nu+\sqrt{2}\,\frac{1-2\nu}{1-\nu}\,\rho^2 }{ 4-5\nu+2\left( 4-\sqrt{2}\,\frac{1-2\nu}{1-\nu} \right)\rho^2 }\,.
\end{equation*}

We wish to show that $0<{\mathcal H}(\rho)<1$. Since the denominator $4-5\nu+2\left( 4-\sqrt{2}\,\frac{1-2\nu}{1-\nu} \right)\rho^2$ does not vanish, ${\mathcal H}$ is a monotone function of $\rho^2$, so that it suffices to consider its extremal values
\begin{equation*}
  {\mathcal H}(0) = \frac{ \nu }{ 4-5\nu } \quad ; \quad {\mathcal H}(+\infty) = \frac{ \sqrt{2}\,\frac{1-2\nu}{1-\nu} }{ 2\left( 4-\sqrt{2}\,\frac{1-2\nu}{1-\nu} \right) }
       = \frac{ \sqrt{2}\,(1-2\nu) }{ 2\left[ 4(1-\nu)-\sqrt{2}\,(1-2\nu) \right] }\,.
\end{equation*}
Both of these quantities are homographic, monotone functions of $\nu$; they vary between the extremal values (obtained for $\nu=0$ and $\nu=1/2$ respectively) $0$ and $1/3$, $\frac{ \sqrt{2} }{ 2(4-\sqrt{2}) } \simeq 0.273$ and $0$, respectively. All of these values are in the interval $(0,1)$, so the same is true of the quantities ${\mathcal H}(0)$ and ${\mathcal H}(+\infty)$, and of ${\mathcal H}(\rho)$ itself for all values of $\rho$.

Now rewrite the above inequality in the form
\begin{equation*}
  \begin{array}{l}
    (1-\nu)\left[2-\nu - (2-3\nu){\mathcal H}(\rho)\right] \\[3mm]
    \ds \quad\quad\quad\quad + \left\{ 2+\nu-2\sqrt{2}\,(1-2\nu) + \left[ 3(2-\nu) - 4\sqrt{2}(1-2\nu) \right] {\mathcal H}(\rho) \right\} \rho^2
     >? \ 0.
  \end{array}
\end{equation*}
Since $0<{\mathcal H}(\rho)<1$, the term $2-\nu - (2-3\nu){\mathcal H}(\rho)$ is larger than $2-\nu-(2-3\nu) = 2\nu$ and the term $2+\nu-2\sqrt{2}\,(1-2\nu) + \left[ 3(2-\nu) - 4\sqrt{2}(1-2\nu) \right] {\mathcal H}(\rho)$ larger than $2+\nu-2\sqrt{2}\,(1-2\nu)$. Hence for this inequality to be true, it suffices that the inequality
\begin{equation*}
    2\nu(1-\nu) + \left[ 2+\nu-2\sqrt{2}\,(1-2\nu) \right] \rho^2 >? \ 0
\end{equation*}
be satisfied. For this to be true for all values of $\rho$, it suffices that
\begin{equation*}
  2+\nu-2\sqrt{2}\,(1-2\nu) >? \ 0 \quad \Leftrightarrow \quad \nu >? \ \frac{2\sqrt{2}-2}{4\sqrt{2}+1} = \frac{2}{31}(9-5\sqrt{2}) \simeq 0.124.
\end{equation*}

{\it This reasonable condition is assumed to be satisfied in the sequel}.\footnote{It has already been noted by \cite{LKL11}, in the case of a fracture energy independent of mode mixity, that the case of small positive values of $\nu$ is special.} Then $\frac{\partial\Phi}{\partial\xi}(\rho;0) > 0$ for $\rho<\rho^{\rm cr}$.

\subsection{Consequences for the equation $\Phi(\rho^0;\xi)=0$ for $\rho^0<\rho^{\rm cr}$}\label{subapp:EqOnXiSmallRho}

Let us assume that $\rho^0<\rho^{\rm cr}$. Then:
\begin{itemize}
  \item At the point $\xi=0$, $\Phi(\rho^0;0)>0$, see Subsection \ref{subapp:StudyPhi}.
  \item At the same point $\frac{\partial\Phi}{\partial\xi}(\rho^0;0) > 0$, see Subsection \ref{subapp:PosDeriv}.
  \item The function $\frac{\partial\Phi}{\partial\xi}(\rho^0;\xi)$ increases with $\xi$, see Subsection \ref{subapp:StudyPhi}; hence $\frac{\partial\Phi}{\partial\xi}(\rho^0;\xi) > 0$ for $\xi>0$.
\end{itemize}
One deduces from these elements that when $\xi$ increases from $0$ to $+\infty$, $\Phi(\rho^0;\xi)$ increases from some positive value; hence the equation $\Phi(\rho^0;\xi)=0$ on $\xi$ does not admit any solution in the interval $(0,+\infty)$.